\documentclass[11pt,a4paper]{article} 
\pdfoutput=1
\usepackage{jheppub,cancel}
\usepackage{mathrsfs}
\DeclareMathAlphabet{\mathpzc}{OT1}{pzc}{m}{it}

\usepackage[english]{babel}

\usepackage{amsmath}
\usepackage{amssymb}

\usepackage{mathrsfs}
\DeclareMathAlphabet{\mathpzc}{OT1}{pzc}{m}{it}

\usepackage{graphicx}

\usepackage{cancel} % Cro
\usepackage{paralist}
\usepackage{booktabs}
\usepackage{hyperref}
\usepackage{xspace}

\usepackage{graphicx,color,verbatim,acronym,epsfig,subfigure}
\usepackage{slashed,skak}
\usepackage{latexsym,amsfonts,amssymb,amsmath,makeidx,mathrsfs,multirow,lineno,bm,bbm}
\newcommand{\Z}{\mathbb{Z}} % Integers

\newcommand{\abs}[1]{|#1|} % Absolute value
\definecolor{nicegreen}{rgb}{0.1,0.7,0.1}

\newcommand{\be}{\begin{equation}}
\newcommand{\ee}{\end{equation}}
\newcommand{\bea}{\begin{eqnarray}}
\newcommand{\eea}{\end{eqnarray}}

\def\beqn{\begin{eqnarray}}
\def\eeqn{\end{eqnarray}}
\def\beqs{\begin{subequations}}
\def\eeqs{\end{subequations}}
\def\beq{\begin{equation}}
\def\eeq{\end{equation}}
\def\ba{\begin{array}}
\def\ea{\end{array}}

\def\[{\left[}
\def\]{\right]}
\def\({\left(}
\def\){\right)}

\def\beq{\begin{equation}}
\def\eeq{\end{\equation}}
\def\bea{\begin{eqnarray}}
\def\eea{\end{eqnarray}}

\def\to{\rightarrow}

\def\nn{\nonumber}

\def\be{\beta}

\def\ifmath#1{\relax\ifmmode #1\else $#1$\fi}
\def\ls#1{\ifmath{_{\lower1.5pt\hbox{$\scriptstyle #1$}}}}
\def\lss#1{\ifmath{^{\,\lower2.5pt\hbox{$\scriptstyle #1$}}}}

\def\beq{\begin{equation}}
\def\eeq{\end{equation}}

\def\lsim{\mathrel{\raise.3ex\hbox{$<$\kern-.75em\lower1ex\hbox{$\sim$}}}}
\def\gsim{\mathrel{\raise.3ex\hbox{$>$\kern-.75em\lower1ex\hbox{$\sim$}}}}

\def\bit{\begin{itemize}}
\def\eit{\end{itemize}}
\def\bec{\begin{center}}
\def\eec{\end{center}}
\def\bed{\begin{description}}
\def\eed{\end{description}}

\usepackage{ulem,fancyvrb}
\usepackage{xcolor}

\begin{document}

\title{ Two-step strongly first-order electroweak phase transition modified FIMP dark matter, gravitational wave signals, and the neutrino mass}
% with the scotogenic model

\author[\alpha\,, \beta]{~Ligong Bian,}
\emailAdd{lgbycl@cqu.edu.cn}
\affiliation[\alpha]{ Department of Physics, Chongqing University, Chongqing 401331, China}
\affiliation[\beta]{ Department of Physics, Chung-Ang University, Seoul 06974, Korea}
\author[\delta]{Xuewen Liu}
\affiliation[\delta]{Department of Physics, Nanjing University, Nanjing, 210093, China}
\emailAdd{xuewenliu@nju.edu.cn}

\date{\today}

\abstract{
We study a dynamical freeze-in production of the dark matter considering the electroweak phase transition history of the Universe. The kinematical thresholds of the decay and scattering processes for the dark matter production can be altered by the temperature dependent thermal masses of particles, which might lead to enhancement or reduction of the dark matter relic abundance. The second-stage strongly first-order electroweak phase transition (SFOEWPT) triggered by the hidden scalars can be probed at colliders and the gravitational wave detectors. The two-step SFOEWPT modified late decay FIMP dark matter is accomplished with a Dirac neutrino mass explanation in the scotogenic model.
}

\maketitle
\preprint{}

%\setcounter{page}{2}

%\tableofcontents

%----------------------------------------------------------------------
%\vspace*{1.5cm}

%\newpage

\section{\hspace*{-2mm}Introduction}
\label{section:introduction}

%=======================================================
The dark matter (DM), as an essential ingredient of the standard cosmological model, is supported by cosmological observations~\cite{Ade:2015xua} 
and can be explained as particles beyond the standard model (SM). 
The non-gravitational property of the DM is not known to us~\cite{Bertone:2016nfn}.  
The explanations of the DM motivates the Weakly Interacting Massive Particles (WIMPs), 
where the DM weakly interacts with the SM sector but strongly enough to be probed at various 
direct detection experiments~\cite{Tan:2016zwf,Akerib:2016vxi,Aprile:2017iyp}, 
indirect detection experiments~\cite{Fermi-LAT:2016uux,Abdallah:2016ygi,Abdalla:2016olq,HESS:2015cda,Aguilar:2016kjl}  
and colliders, such as LHC, and future CEPC, ILC, and FCC-ee. 
However, the lack of the signals of WIMPs at these detectors motivates people to go beyond the WIMPs paradigm~\cite{Baer:2014eja}. 
In the WIMPs paradigm, the DM particles live in the thermal bath at high temperature through the interaction with the SM particles, 
and the DM relic abundance is produced by the freeze-out mechanism after the DM particles annihilation rate falls below 
the Universe expansion rate at temperature around $T_{fo}\sim m_{DM}/x_{fo}$ with $x_{fo}\sim 26$~\cite{ Arcadi:2017kky}. 
One alternative approach can be the Feebly Interacting Massive Particle (FIMP), 
where the DM particles never enter into thermal equilibrium with the SM plasma due to the super-weak or 
feebly interaction rate between the DM particles and the SM particles~\cite{McDonald:2001vt,Hall:2009bx,Bernal:2017kxu}. 
In this paradigm, a basic assumption is that the initial DM abundance (at reheating epoch) is negligible\footnote{The population of DM for the initial negligible DM abundance is related to the deep understanding of the early universe.
Due to the FIMP DM never enter into the thermal equilibrium, primordial perturbations in the DM
density spectrum would not be washed out and may leave an imprint
in the CMB, especially when the FIMP DM is the scalar inflaton~\cite{Bernal:2017kxu,Tenkanen:2016jic}. 
This feature is supposed to limit the FIMP models~\cite{Bernal:2017kxu}. 
The interaction strength between the inflaton and the DM sector might be bounded by 
the DM relic density~\cite{Adshead:2016xxj,Kofman:1994rk,Shaposhnikov:2006xi,Bezrukov:2008ut,Dev:2013yza}.
We left the connection of this work with the inflation and the reheating to a separate publication.}, and the present DM abundance is produced by the so-called freeze-in mechanism, 
through the SM particles (in the thermal bath) decaying or annihilating to the DM particles 
at high temperature around $T_{fi}\sim m_{DM}/x_{fi}$ with $x_{fi}\sim 1$~\cite{Hall:2009bx,Blennow:2013jba,Bernal:2017kxu}. 
There also exist some other non-thermal DM production mechanisms such as incomplete reheating~\cite{Kofman:1994rk}, asymmetric reheating~\cite{Hodges:1993yb,Berezhiani:1995am,Adshead:2016xxj}, and Dodelson-Widrow mechanism~\cite{Dodelson:1993je}, however, we do not consider in this work.

In the FIMP scenario, the $T_{fi}\sim\mathcal{O}(10^2)$ GeV for the weak scale DM, 
this temperature is in the ballpark of the electroweak phase transition temperature of the Universe. 
One can expect that the thermal effects from the electroweak phase transition change the production of the FIMP DM
by modifying the kinematical threshold of the DM decay and annihilation processes, 
recent studies can be found in Refs.~\cite{Baker:2016xzo,Baker:2017zwx,Bian:2018mkl}
\footnote{For the phase transition modified WIMPs scenario, we refer to Ref.~\cite{Hektor:2018esx}.}.
The electroweak phase transition provides an explanation of the electroweak symmetry breaking of the SM at the early Universe, 
which can be tested by probing the triple Higgs couplings at high energy hadronic colliders~\cite{Arkani-Hamed:2015vfh} 
and the future linear colliders, such as CEPC, ILC, and FCC-ee. 
On the other hand, supposing the electroweak phase transition is a strongly first order phase transition (SFOEWPT), 
the space-based gravitational wave detectors can probe the gravitational wave signals from the phase transition.

Recently, the hidden sector triggered multi-step phase transition arouse peoples interests, see Ref.~\cite{Patel:2012pi,Cheng:2018ajh,Curtin:2014jma,Chao:2017vrq} for the combined studies of the WIMP DM and the two-step SFOEWPT. Wherein, the first-step is a second-order phase transition, and the second-step is a SFOEWPT. There also exist a broad class of well-motivated DM models with more than one hidden sectors, wherein the DM particle is one component of the hidden sectors~\cite{Pospelov:2007mp,Berlin:2016vnh}. In this work, we study the two-step SFOEWPT scenario with the FIMP DM particle being one component of the hidden sectors. 
The non-zero neutrino mass is well established by the oscillation experiments, which requires for new physics beyond the SM~\cite{deSalas:2018bym}.\footnote{The bounds on the summation of the neutrino masses from the cosmology are given recently by Ref.~\cite{Aghanim:2018eyx}. }
Therefore, it would be inspiring if the neutrino mass can be addressed together with the two-step SFOEWPT modified FIMP DM with the DM being a part of hidden sectors.
This constitute the essential ingredient of this work. The scotogenic model proposed by
E. Ma~\cite{Farzan:2012sa} falls into the context of the DM models, where the Dirac neutrino mass is radiatively generated with the agumented of two hidden sectors (including an inert scalar doublet and a scalar singlet).\footnote{For the study on WIMP DM as the mixing of the two scalars we refer to Ref.~\cite{Liu:2017gfg} at the one-loop level.} The DM particles in the model can be the lightest of the heavy neutrinos or the neutral scalars from the dark sector.\footnote{For relevant studies within the WIMP paradigm can be found in Ref~\cite{Farzan:2012sa} and the references therein.
Ref.~\cite{Molinaro:2014lfa} studied the FIMP realization of DM within the original scotogenic model of Ref.~\cite{Ma:2006km}, where the Majorana neutrino mass is radiatively generated by the help of inert scalar sector. }

This work is organized as follows: We firstly review the scotogenic model in Section~\ref{S:SIDM}. The phase transition patterns and the implications for the FIMP DM production are explored in Section~\ref{sec:PTdyn}. The DM phenomenology is studied in Section~\ref{sec:DMP}.
The gravitational wave signals predictions and related collider interplay are discussed in Section~\ref{sec:GWs} and Section~\ref{sec:cols}. We conclude with Section~\ref{sec:conl}.

%=======================================================

\section{The scotogenic model }
\label{S:SIDM}

%%%%%%%%%%%%%%%%%%%%

We revisit the scotogenic model that generates the Dirac neutrino mass with dark matter at one-loop level~\cite{Farzan:2012sa}, the Yukawa interactions for the radiative Dirac neutrino mass generation are given by
\begin{align}\label{eq:yuk}
\mathcal{L}_Y=f_{\alpha\kappa} \bar{N}_\kappa (\frac{1-\gamma_5}{2})(\nu_\alpha H_0-l_\alpha H^+) 
+ h_{\kappa\beta}\bar{N_\kappa}(\frac{1+\gamma_5}{2})\nu_\beta S +\mathrm{h.c.}, 
\end{align}
where the introduced inert doublet ($\eta$) together with an additional real singlet scalar ($S$) are all odd under  
the $\Z_2$ symmetry, the tree level potential of the model is 
\bea\label{eq:tre}
V &=& \mu_{\Phi}^2 {\Phi}^\dagger {\Phi} + \mu_{\eta}^2  {\eta}^\dagger {\eta} + 
\frac{\mu_S^2}{2} S^2 + \lambda_1 ({\Phi}^\dagger {\Phi})^2
+ \lambda_2 ({\eta}^\dagger {\eta})^2 + \lambda_3
(\Phi^\dagger \Phi)(\eta^\dagger \eta) \nonumber \\ &+&
\lambda_4 (\eta^\dagger \Phi)(\Phi^\dagger \eta) + {1 \over 2} \lambda_5
[(\eta^\dagger \Phi)^2 + (\Phi^\dagger \eta)^2] + \frac{\lambda_s}{4} S^4 +  \lambda_{s\phi} 
S^2 (\Phi^\dagger \Phi) \nonumber\\&+&  \lambda_{s\eta}
S^2 (\eta^\dagger \eta)+\mu_{\text{soft}}  S (\Phi^\dagger \eta + \eta^\dagger \Phi)\, .
\eea  
%assuming all the couplings ($\lambda_i,\,\, i=1,8$) in Eq.~\ref{eq:tre} are real. 
In the global minimum of the Electroweak (EW) vacuum,  the doublets are given as
\bea
\Phi = \left( \begin{array}{c}
                           G^+  \\
        \frac{1}{\sqrt{2}}(v+h+i G^0)  
                 \end{array}  \right) \, ,                     
&& \eta =\left( \begin{array}{c}
                           H^+   \\
        \frac{1}{\sqrt{2}}(H_0+iA)  
                 \end{array}  \right) \, ,
\label{eq:fd}
\eea 
where $\Phi$ develops a VEV $v=246$ GeV. The $\mathbb{Z}_2$ symmetry of $\eta, S$ 
remains unbroken in the EW vacuum. Goldstones $G^+$ and $G^0$ are eaten by $W^\pm$, $Z$ bosons after 
the spontaneous symmetry breaking (SSB).  
We consider CP conserving situation in this work and all soft masses and scalar quartic couplings in Eq.~\ref{eq:tre} are assumed to be real,  the parameter relations and relevant theoretical and experimental constraints are given in Appendix~\ref{sec:cons}. In Ref.\cite{Babu:2007sm}, a one-step SFOEWPT is reachable due to a sizable dimensional-six operator of Higgs after integrating out the heavy scalars in the scalar potential. While in this study we focus on the two-step phase transition since the dimensional-six operator can only be obtained at loop level, see Ref.~\cite{Cheng:2018axr,Blinov:2015vma}, which is different from our previous study on mixed scalar WIMPs DM~\cite{Liu:2017gfg} where a sizable interaction rate between the SM Higgs sector and the dark scalars are allowed.

In the EW vacuum, the mixing between the singlet and the inert doublet neutral scalars $S,H_0$ connects the gauge eigenstates to the mass 
eigenstates $\chi$ and $H$ by
\begin{align}
&\quad\left(\begin{array}{c}
    S \\
    H_0 \\
  \end{array}\right)
=\left(  \begin{array}{cc}
    \cos\theta & -\sin\theta \\
    \sin\theta & ~\cos\theta \\
  \end{array}\right)
\left(\begin{array}{c}
    \chi \\
    H \\
  \end{array}\right)\;.
\end{align}
Here, the mixing angle is
\begin{align}
\theta=\frac{1}{2}\tan^{-1}(\frac{2v \mu_{\text{soft}}}{\tilde{M}_S^2-\tilde{M}_{H_0}^2})\;,
\end{align}
with $\tilde{M}_S^2 =2 \mu_S^2 + \lambda_{s\phi}  v^2 $ and
$\tilde{M}_{H_0}^2 = \mu_\eta^2 +  \lambda_{L}v^2 $. For parameter relations between mass eigenstates and gauge eigenstates, see Appendix~\ref{sec:pars}. 

 \begin{figure}[!htp]
\begin{centering}
\includegraphics[width=.3\textwidth]{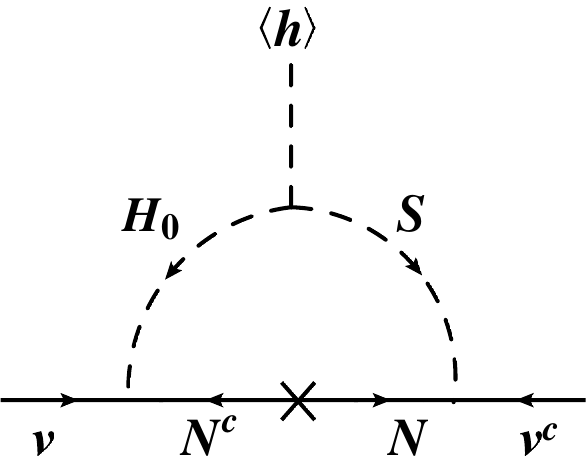}
\caption{The neutrino mass generation in the scotogenic model.} \label{fig:MN}
\end{centering}
\end{figure}

From  Eq.~\ref{eq:yuk}, one gets the one-loop Dirac neutrino mass,
 \begin{eqnarray}\label{eq:neumM}
\left(M_\nu\right)_{\alpha\beta}=\frac{\sin2\theta}{32\sqrt{2}\pi^2}\sum_\kappa f_{\alpha \kappa}h_{\kappa\beta} m_{N_\kappa}\Big[\frac{m_\chi^2}{m_\chi^2-m_{N_\kappa}^2}\log \frac{m_\chi^2}{m_{N_\kappa}^2}- \frac{m_H^2}{m_H^2-m_{N_\kappa}^2}\log \frac{m_H^2}{m_{N_\kappa}^2}\Big]\;,
\end{eqnarray}
with the Feynman diagram given in the Fig.~\ref{fig:MN}. 

For the scotogenic model proposed in Ref.~\cite{Ma:2006km}, the dangerous of flavor-changing charged-lepton radiative decays can be avoided by assuming a superweak 
Yukawa interaction of $f_{\alpha\kappa}$, which prefers a FIMP DM~\cite{Hall:2009bx,Molinaro:2014lfa,Hessler:2016kwm}. Indeed, a superweak interaction of $N_1$ with scalars leads to the decouple of $N_1$ from the neutrino mass matrix, and there are only two light neutrinos acquire non-zero masses, see Ref.~\cite{Molinaro:2014lfa}.
For the Dirac neutrino mass generation in the scotogenic model proposed by Ref.~\cite{Farzan:2012sa}, the extra scalar are introduced to accomodate the second yukawa interaction of the Eq.~\ref{eq:yuk}. There is no problem of the flavor-changing charged-lepton radiative decays. In this work, we would explore the $N_1$ as the FIMP DM, in which case the $N_1$ decouples from the neutrino mass matrix.  
Furthermore, the smallness of the neutrinos masses can be ensured by the smallness mixing between 
$S$ and $H_0$ in the model, which naturally leads to the production of $\chi$ through freeze-in mechanism, which then decay to the DM particle $N_1$ after the freeze-in.   
This study applies to the Global $U(1)_{B-L}$ scenario of Ref.~\cite{Farzan:2012sa} and the millicharge situation of the gauged $U(1)_{B-L}$ scenario.

\section{The electroweak phase transition dynamics}
\label{sec:PTdyn}

Supposing $H,A, H^\pm$ and $S$ are all much heavier than the SM-like Higgs, one can safely integrate out the dark scalars, 
and therefore the tree-level potential with dimensional-six operators is given by  
\begin{eqnarray}
V_0=\mu_\Phi^2 |\Phi|^2+\lambda |\Phi|^4+c_6 |\Phi|^6\;,
\end{eqnarray}
with the $c_6$ mainly coming from the one-loop level, given by Ref.~\cite{Cheng:2018axr,Blinov:2015vma} respectively for the dark scalar and inert doublet cases. Therefore, to obtain a strongly first order phase transition with one-step pattern, a relatively larger quartic coupling is required~\cite{Cheng:2018axr}. This may result in perturbativity problem~\cite{Cheng:2018axr,Cheng:2018ajh,Curtin:2014jma,Blinov:2015vma} even at the model phenomenological valid energy scales. Another reason that motivates us to focus on the two-step pattern phase transition is to account for the production of scalar $\chi$ and the DM $N_1$ through freeze-in mechanism.

\subsection{Electroweak phase transition types and vacuum structures}

\begin{figure}[!htbp]
\begin{centering}
\includegraphics[width=.4\textwidth]{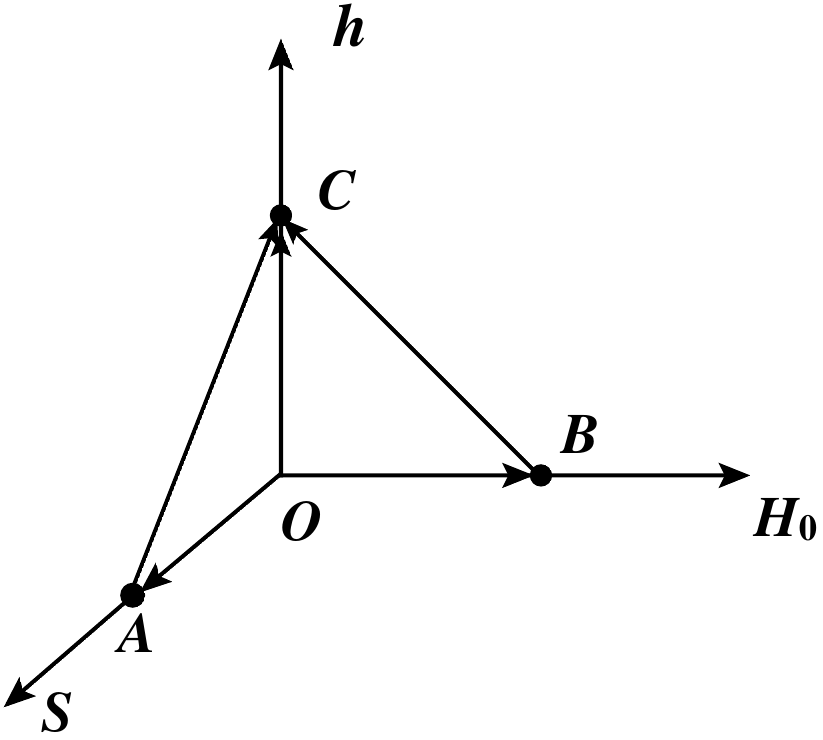}
\caption{The phase transition patterns.} \label{fig:ptt}
\end{centering}
\end{figure}

Firstly, we analyse the vacuum structure at zero temperature, which is highly related with the possible phase transition types, i.e., one-step and two-step. The tree-level scalar potential of the classical fields is given by
\begin{eqnarray}\label{eq:trz}
V_0(h,H_0,S)&=&\frac{\mu_\Phi^2}{2}h^2
+\frac{\mu_\eta^2}{2}H_0^2+\frac{\mu_S^2}{2}S^2
+\frac{\lambda_1}{4}h^4+\frac{1}{4}(\lambda_3+\lambda_4+\lambda_5)h^2H_0^2+\frac{\lambda_2}{4}H_0^4\nonumber\\
&&+\frac{1}{2}\lambda_{s\phi} h^2S^2+\frac{1}{2}\lambda_{s\eta}H_0^2 S^2+\frac{\lambda_s}{4}S^4 + \mu_{soft}hH_0S\;.
\end{eqnarray}
In the space of $(h,H_0,S)$, the minimization conditions,
\begin{eqnarray}
\frac{\partial }{\partial h}V_0(h,H_0,S)=0\;, ~\frac{\partial }{\partial H_0}V_0(h,H_0,S)=0\;, ~\frac{\partial }{\partial S}V_0(h,H_0,S)=0\;, 
\end{eqnarray}
give rise to the possible minima of the potential localized at
\begin{eqnarray}
\label{eq:vacst}
&&(\pm\sqrt{-\mu_\Phi^2/\lambda_1},0,0),~(0,\pm\sqrt{-\mu_\eta^2/\lambda_2},0),~(0,0,\pm\sqrt{-\mu_S^2/\lambda_s}),\\&&(0,\pm\sqrt{(\mu_S^2\lambda_{s\eta}-\mu_I^2\lambda_s)/(\lambda_2\lambda_2-\lambda_{s\eta}^2)},\sqrt{(\mu_\eta^2\lambda_{s\eta}-\mu_s^2\lambda_2)/(\lambda_2\lambda_2-\lambda_{s\eta}^2)})\;.
\end{eqnarray}
We note that the possible minima where h and S(or/and $H_0$) have VEVs is precluded because both the inert doublet and the real singlet scalar are all odd under the $\Z_2$  symmetry.
The EW vacuum localized at $(\pm\sqrt{-\mu_\Phi^2/\lambda_1},0,0)$ should be the global minimum, which ensures the possibility of the phase transition type of $O\to C$ in Fig.~\ref{fig:ptt}.
There also could be three other local minima in the direction of $(0,\langle H_0\rangle,0)$, $(0,0,\langle S\rangle)$  for the case where the mixing of $H_0,S$ is negligible, and in the subspace of $(0,\langle H_0\rangle,\langle S\rangle)$ when the mixing of $H_0,S$ is non-negligible which however is not favored by the neutrino mass generation. To ensure the EW vacuum to be the global vacuum, the following conditions need to be fulfilled
\begin{eqnarray}
\Delta V_0(h,S)\equiv V_0(0,0,S)-V_0(h,0,0)>0\;,~
\Delta V_0(h,H_0)\equiv V_0(0,H_0,0)-V_0(h,0,0)>0\;,
\end{eqnarray}  
which result in the following two bounds on the parameter spaces,
\begin{eqnarray}\label{eq:vachH}
\lambda_s\mu_\Phi^4>\lambda_1\mu_s^4\;, ~\lambda_2\mu_\Phi^4>\lambda_1\mu_\eta^4\;.
\end{eqnarray}

In the spirit of the gauge invariant~\cite{Patel:2011th,Chao:2017vrq}, the finite-temperature potential can be
described by Eq.~\ref{eq:trz} with the substitutions of $ \mu_i^2\rightarrow \mu_i^2(T)$ where
\begin{align}
			\mu_\Phi^2(T) = \mu_\Phi^2 + c_\Phi T^2~,~~
			c_\Phi&=\frac{6\lambda_1 +2\lambda_3 + \lambda_4+\lambda_{s\phi}}{12} + \frac{3 g^2 + g^{\prime2}}{16} + \frac{y_t^2}{4}\\ 
			\mu_\eta^2(T) = \mu_\eta^2 + c_\eta T^2~,~~ 
			c_\eta&=\frac{6\lambda_2 +2\lambda_3 +\lambda_4+\lambda_{s\eta}}{12}+\frac{3g^2 + g^{\prime2}}{16}\\
			\mu_S^2(T) = \mu_S^2 + c_S T^2~,~~ 
			c_S&=\frac{\lambda_S }{4}+\frac{\lambda_{s\phi}+\lambda_{s\eta}}{12}\;.\label{eq:thmu}
	\end{align}
Here, we neglected all other Yukawa couplings apart from the top quark Yukawa coupling $y_t$.
All three directions of $h,H_0,S$ can induce minima due to the thermal corrections at finite temperature depending on these scalar quartic couplings.\footnote{
If one obtains the minimum along the direction of $S$ at finite temperature, the last term of Eq.~\ref{eq:tre} would induce the term of $\mu_{soft}\langle S\rangle _T (\Phi^\dagger \eta+\eta^\dag \Phi)$.  }

The two-step phase transition can occur through $O\to B\to C$ (see Fig.~\ref{fig:ptt}) when the vacuum along the direction of $H_0$ appears earlier than that of the $h$ direction, which corresponds to the parameter constraints of 
\begin{eqnarray}
\label{eq:hHtem}
c_\Phi>0\,,~ c_\eta>0\,,~\mu_\Phi<0\,,~\mu_\eta<0\,,~\mu_\Phi^2 c_\eta>\mu_\eta^2 c_\Phi\;,
\end{eqnarray} 
or occurs through $O\to A\to C$ (see Fig.~\ref{fig:ptt}) when the $Z_2$ symmetry in the direction of $S$ breaks earlier than the SSB of EW
vacuum in the direction of $h$, which results in,
\begin{eqnarray}\label{eq:hstem}
c_\Phi>0\,,~c_S>0\,,~\mu_\Phi^2<0\,,~\mu_S^2<0\,,~\mu_\Phi^2 c_S>\mu_S^2 c_\Phi\;,
\end{eqnarray}

\begin{figure}[!htbp]
\begin{centering}
\includegraphics[width=.4\textwidth]{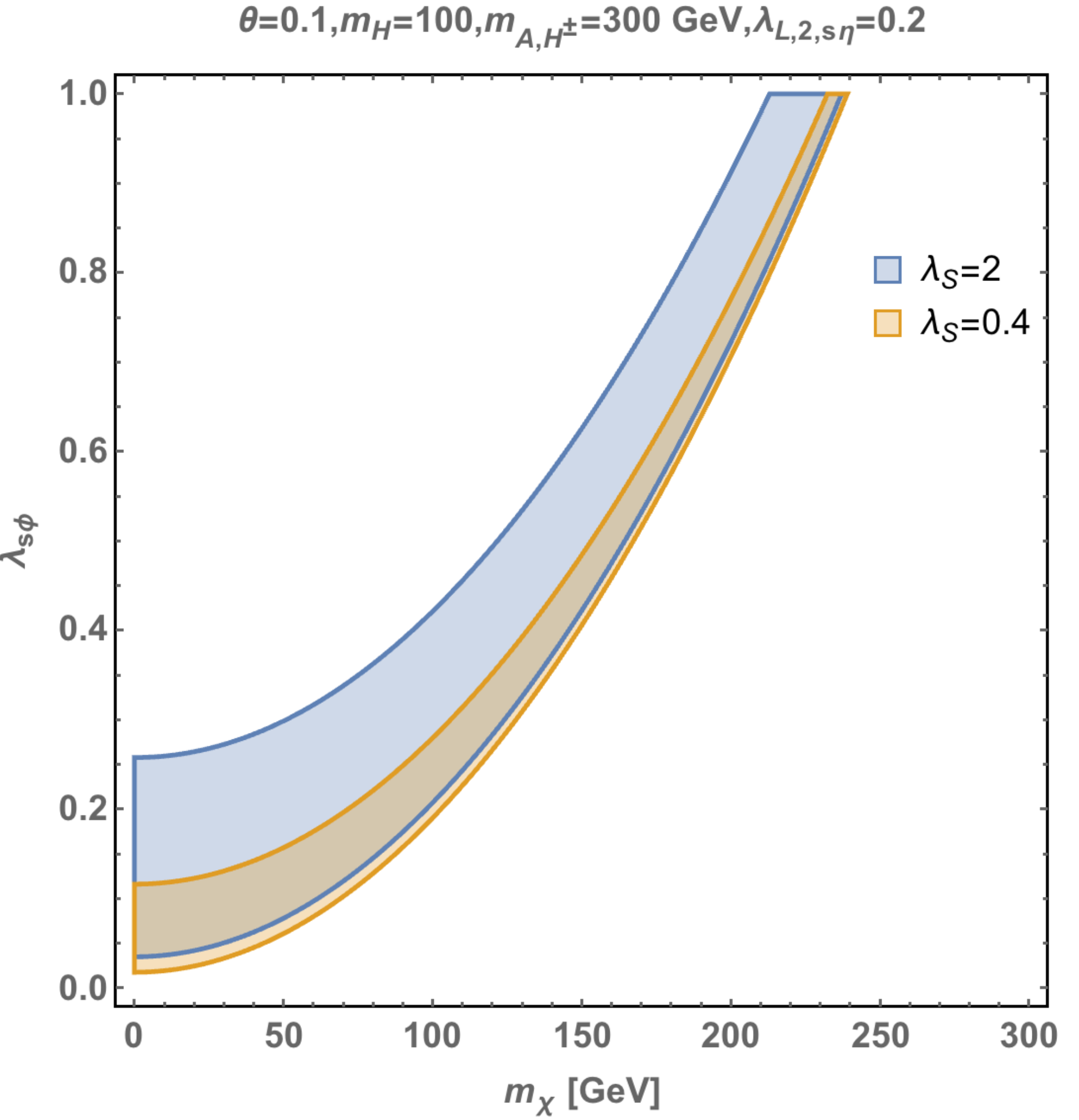}
\includegraphics[width=.4\textwidth]{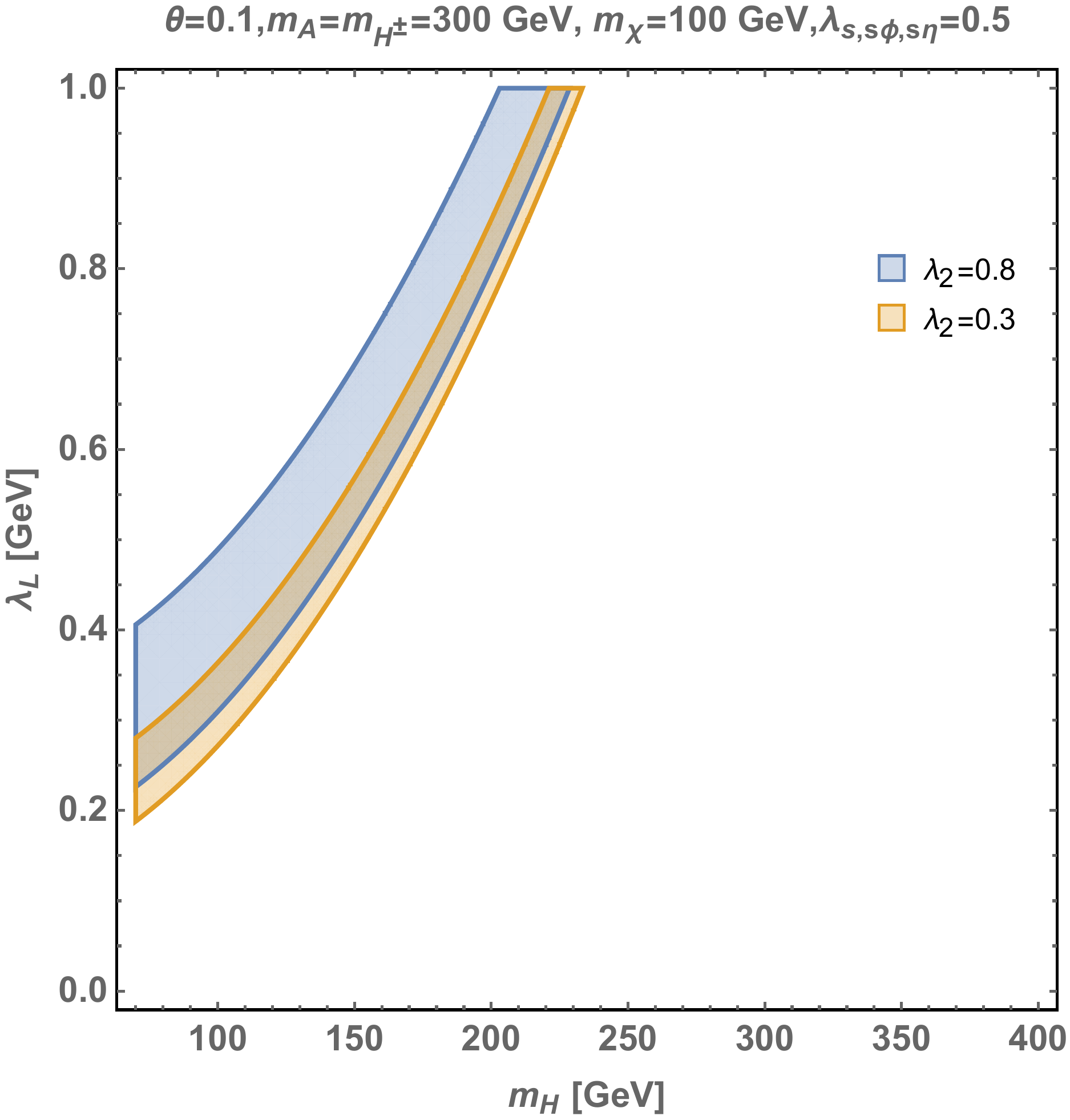}
\caption{Two-step patterns of $O\to A \to C$ (left) and $O \to B \to C$ (right).} \label{fig:idmsinptc}
\end{centering}
\end{figure}

In this work, we study the above three patterns of phase transition. For the one-step pattern, the phase transition can occur directly from the EW symmetry and $Z_2$ symmetry phase to the EW symmetry broken phase with $Z_2$ symmetry.
 For the two-step pattern of $O\to A\to C$, the parameter spaces are restricted by Eqs.~(\ref{eq:vachH},~\ref{eq:hstem}), see the left plot of Fig.~\ref{fig:idmsinptc}.
 For the two-step pattern $O\to B\to C$, Eqs.~(\ref{eq:vachH},\ref{eq:hHtem}) restrict the parameters as shown in the right panel Fig.~\ref{fig:idmsinptc}.

The bounce configuration of the nucleation bubble ( the bounce configuration of the multi-fields that connects the EW broken vacuum ($h-$ vacuum, true vacuum) and the false vacuum( here it can be $Z_2$ broken vacuum $H_0(S)-$ vacuum for the two-step scenarios)) can be obtained by extremizing  
\begin{eqnarray}
S_3(T)=\int 4\pi r^2d r\bigg[\frac{1}{2}\big(\frac{d \phi_b}{dr}\big)^2+V(\phi_b,T)\bigg]\;,
\end{eqnarray}
 through solving the equation of motion for $\phi_b$ (it is $h$ and $H_0/S$ for two-step scenarios),
\begin{eqnarray}
\frac{d^2\phi_b}{dr^2}+\frac{2}{r}\frac{d\phi_b}{dr}-\frac{\partial V(\phi_b)}{\partial \phi_b}=0\;,
\end{eqnarray}
with the boundary conditions of 
\begin{eqnarray}
\lim_{r\rightarrow \infty}\phi_b =0\;, \frac{d\phi_b}{d r}|_{r=0}=0\;.
\end{eqnarray}
The phase transition completes at the nucleation temperature when the thermal tunnelling probability for bubble nucleation per horizon volume and per horizon time is of order unity~\cite{Affleck:1980ac,Linde:1981zj,Linde:1980tt}, 
\begin{eqnarray}
\Gamma\approx A(T)e^{-S_3/T}\sim 1\;.
\end{eqnarray}
 
 \subsection{Thermal mass versus temperature}

\begin{figure}[!htbp]
\begin{centering}
\includegraphics[width=.5\textwidth]{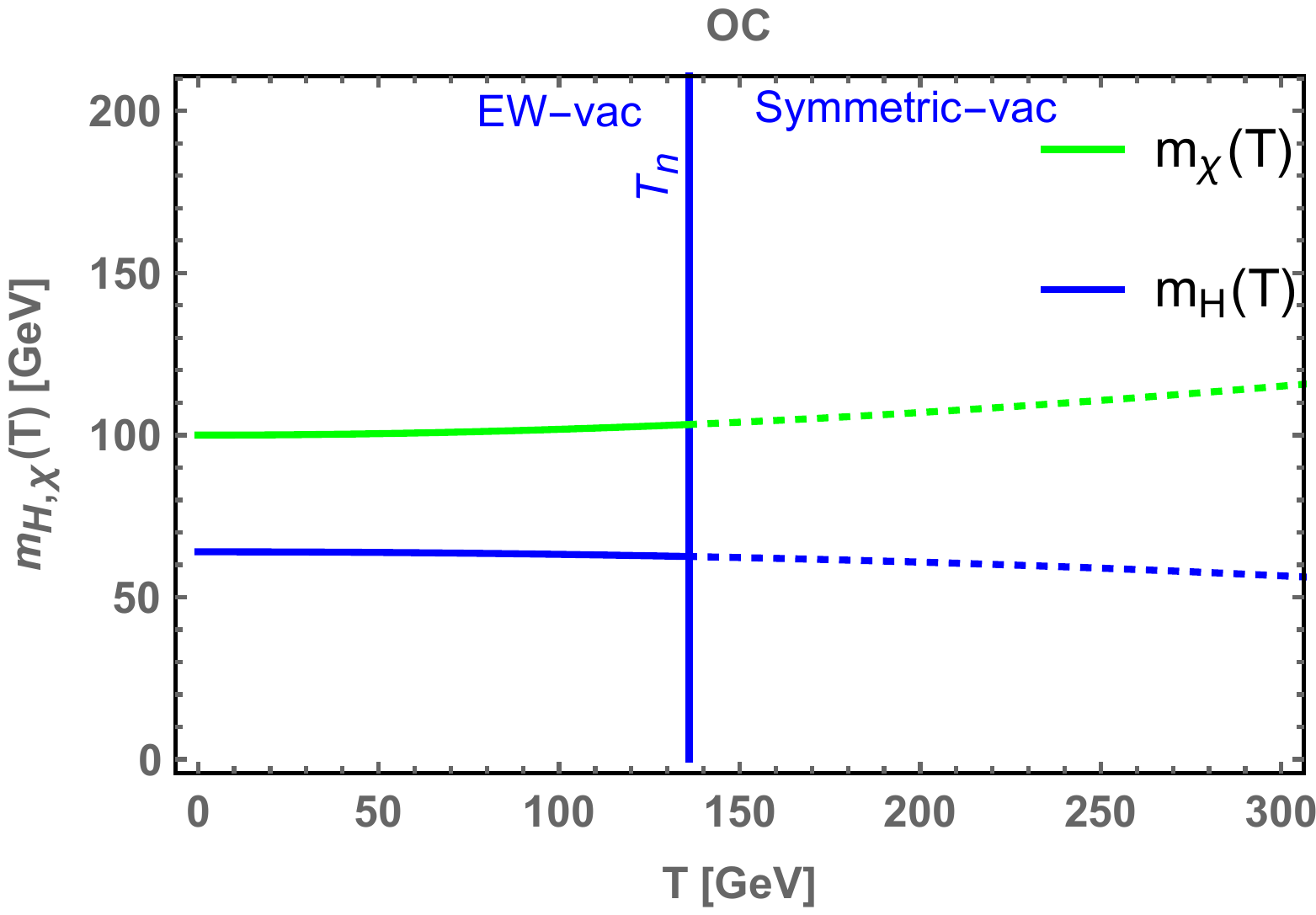}
\caption{Thermal masses for the pattern $O\to C$.} \label{fig:thmOC}
\end{centering}
\end{figure}

\begin{figure}[!htbp]
\begin{centering}
\includegraphics[width=.425\textwidth]{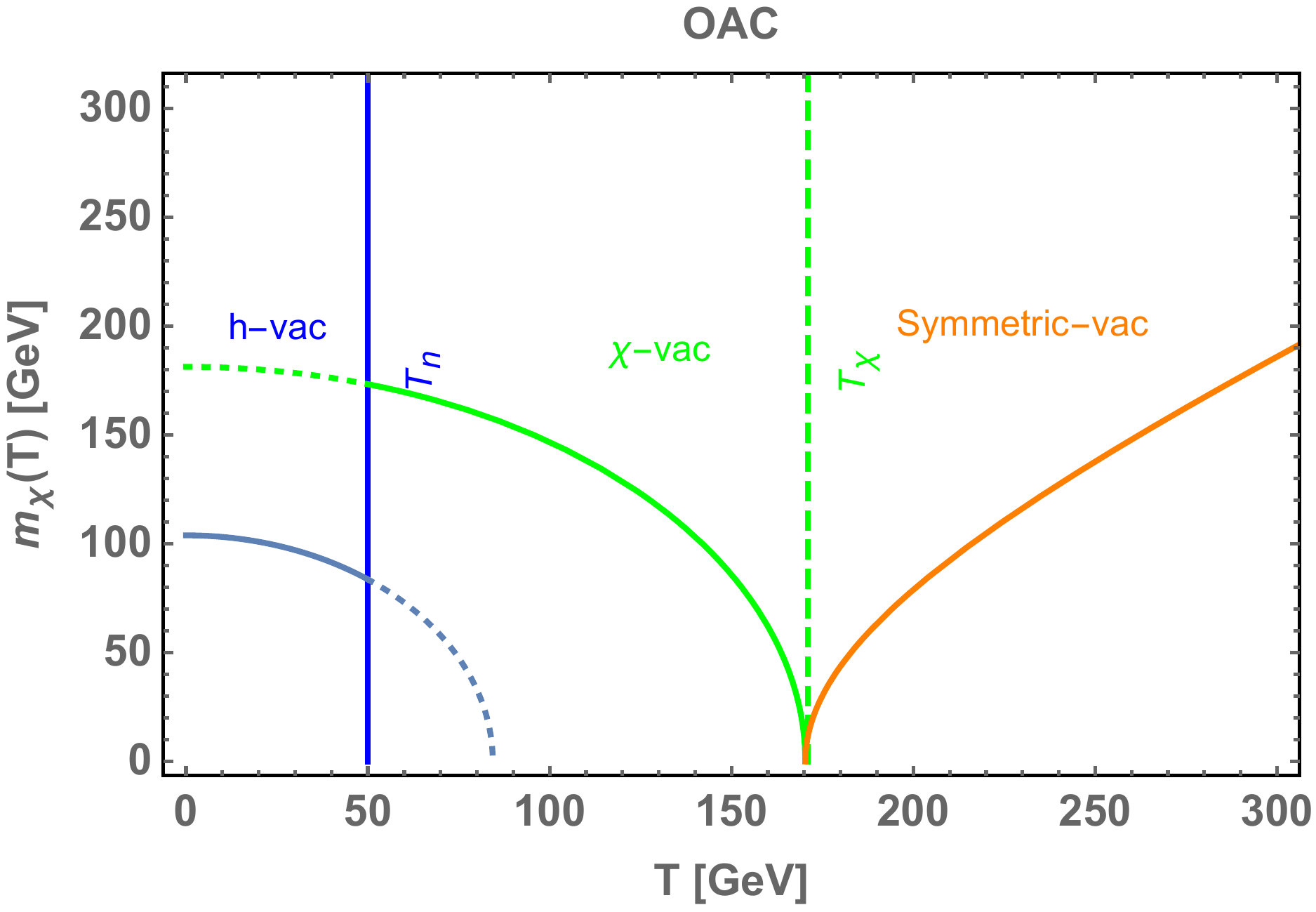}
\includegraphics[width=.425\textwidth]{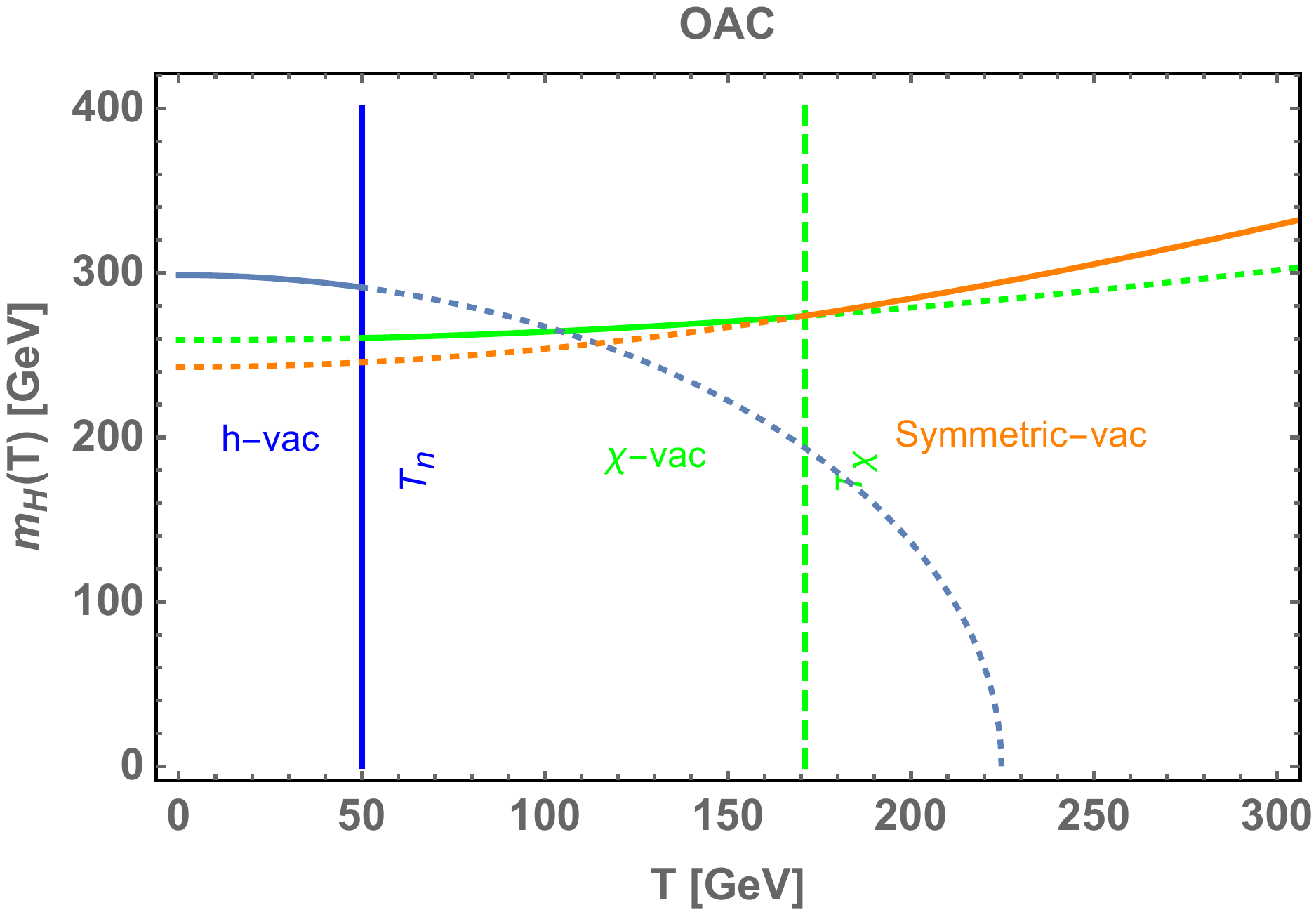}
\caption{Thermal masses for the phase transition pattern $O\to A\to C$.} \label{fig:thmOAC}
\end{centering}
\end{figure}

\begin{figure}[!htbp]
\begin{centering}
\includegraphics[width=.325\textwidth]{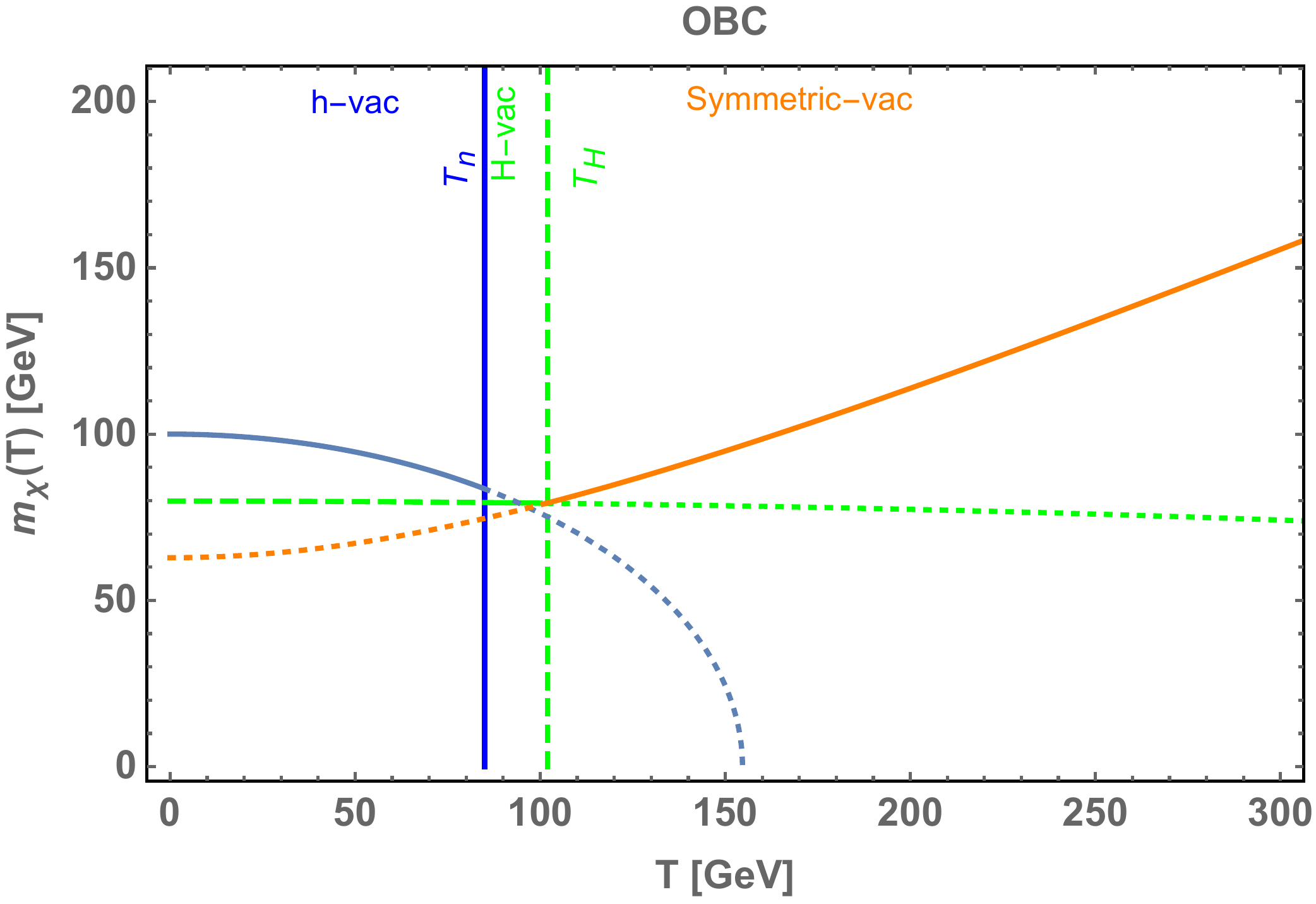}
\includegraphics[width=.325\textwidth]{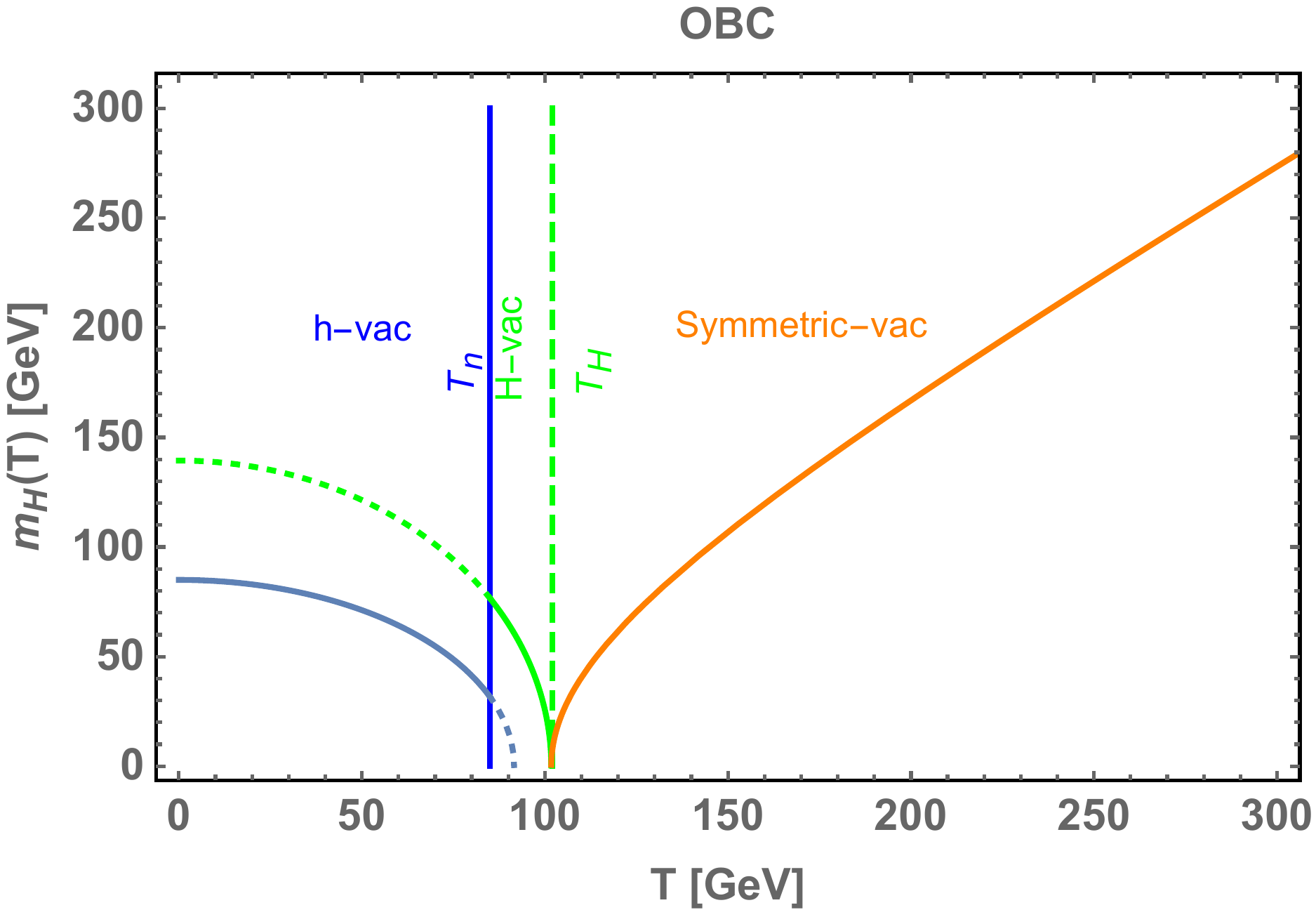}
\includegraphics[width=.325\textwidth]{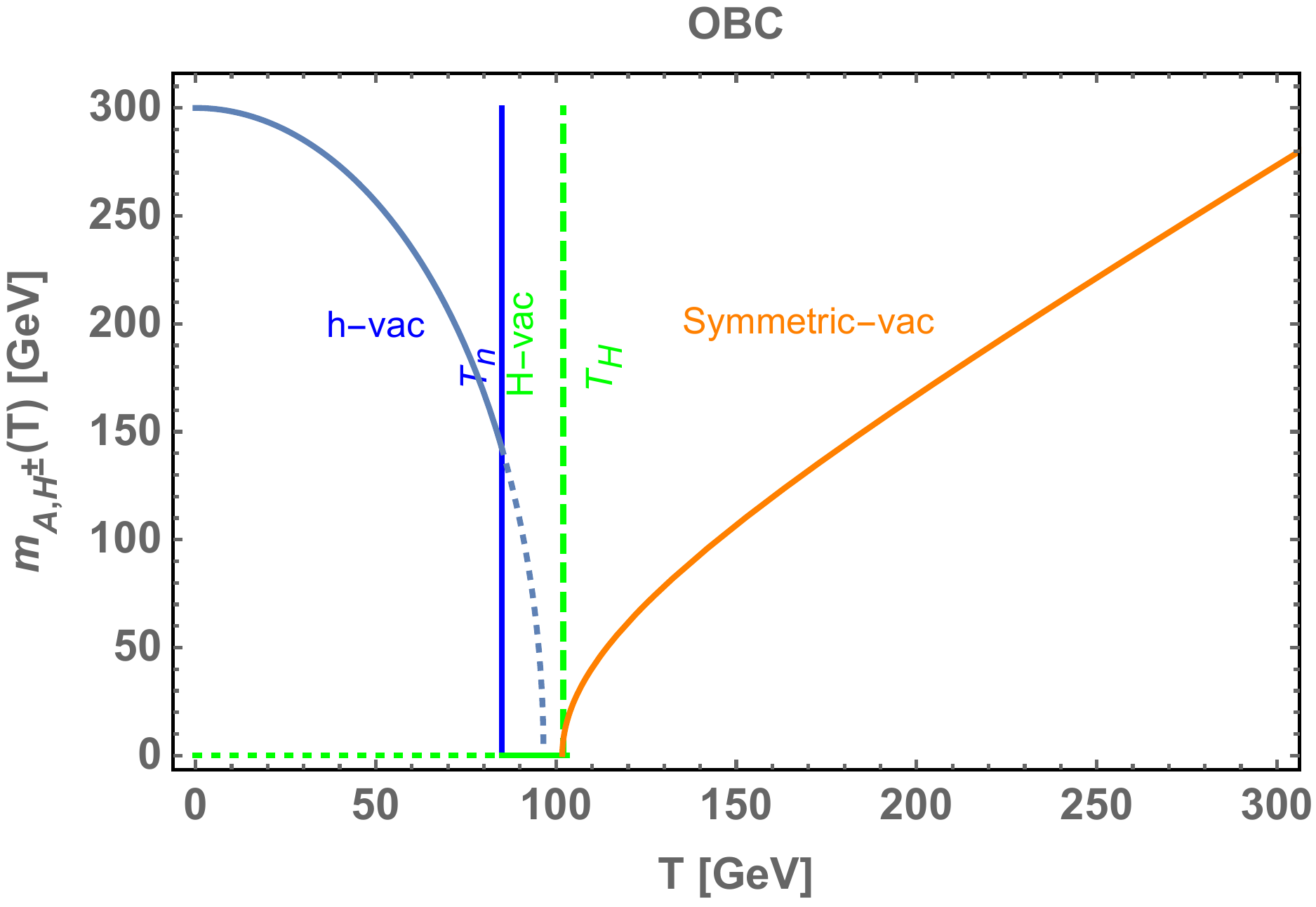}
\caption{Thermal masses for the phase transition pattern $O\to B\to C$.} \label{fig:thmOBC}
\end{centering}
\end{figure}

Before the study of DM dynamics, we first explore the scalar masses evolving with temperature. The thermal masses (see Appendix~\ref{app:thermalmass} for details) are crucial for the thermal decay width and scattering cross section which determine the amplitude and possible evolution history of the DM relic abundance. From Fig.~\ref{fig:thmOC} (the parameters are chosen as in Fig.~\ref{fig:GWmix}), one can find that in the phase transition scenario of $O\to C$, the thermal masses $m_{H,\chi}(T)$ almost equal to the physical masses at $T=0$, therefore, we do not expect a larger deviation of the thermal modified FIMP from the case without taking into account the thermal effects. When  the universe goes through the first-stage second-order phase transition and the second-stage first-order phase transition with the dropping of the temperature, $m_{\chi}(T)$ and $m_{H,H^\pm}(T)$ show a highly dependence on the temperature for the $O\to A\to C$ and $O\to B\to C$ phase transition scenarios, see Fig.~\ref{fig:thmOAC} (with parameters are chosen as the top-right panel of Fig.~\ref{fig:GWsing1} to ensure all dark scalars live in thermal bath during the production process of the DM $N_1$ through freeze-in mechanism) and Fig.~\ref{fig:thmOBC} (with parameters are chosen as the bottom-right panel of Fig.~\ref{fig:GWsing1} and assuming a negligible small $\theta$). Therefore one can expect the kinematic threshold of the decay/inverse decay and the scattering processes that contribute to the DM production can be different from the traditional FIMP DM calculations. More specifically, the mass splitting between $m_{\chi,A,H,H^\pm}(T)$ and $m_{N_1}$ is dynamical and can be vastly different from the zero temperature case in a long time duration, which may lead to a enormous difference of DM relic abundance between thermal modified calculation and the traditional calculation without taking into account the thermal mass. For the case of $\chi$ producing through the freeze-in mechanism, the $\lambda_{s\phi,s\eta}$ should also be negligible, with $m_\chi(T)\approx m_\chi$.

\section{DM phenomenology}
\label{sec:DMP}

For the DM relic density calculation, the accumulated relic abundance of the previous temperature duration will be taken as the initial abundance of the next temperature duration. The DM relic abundance accumulated in the the symmetry phase can be significant depends on the reheating temperature which is taken as the initial condition of the calculation of FIMP where the entropy normalized number density is assumed to be null. For the DM number density calculation with phase transition, we 
use the thermal corrected mass to replace the physical mass at zero temperature in the decay and scattering processes.

\subsection{FIMP $N_1$ DM }
\label{subsec:N1}

In the following, we study the case of $m_{\chi,H,H^\pm,A}>m_{N_1}$ without mass degeneracy.
The relevant study can be found in Ref.~\cite{Molinaro:2014lfa}, here the difference is that we have additional decay of $\chi \rightarrow N\nu$.  As in Ref.~\cite{Molinaro:2014lfa}, the $2\leftrightarrow 2$ process is highly suppressed  in the situation with phase transition pattern $O\to A\to C$, and we therefore focus on the decay/inverse decay of $\chi\to N\nu$ dominated FIMP DM production.
We first study the freeze-in contribution to $N_1$ production, its production will be dominated by the decays of the scalars ($H,A,H^\pm,\chi$) while they are  in equilibrium with the thermal bath. The $N_1$ yield for the $O\to A\to C$ scenario, $Y_{N_{1}}(T)=n_{N_1}(T)/s(T)$, can be computed by solving 
the following Boltzmann equation \cite{Hall:2009bx}

\begin{equation}
\frac{dY_{N}}{dx}\hspace{-0.1cm}=\hspace{-0.1cm}
\frac{1}{sxH}\left(\frac{g_X m_X^2m_N\Gamma_{X\to N\nu/\ell}}{2\pi^2x}\right)K_1\left(\frac{m_X}{m_N}x\right)\;,\label{eq:decayDM}
\end{equation}
where $s$ is the entropy density of the Universe, $H(T)$ is the expansion rate of the Universe at a given temperature, 
with $X=H,A,H^\pm,\chi$ and $\ell$ being a SM lepton. In this equation, $K_{1}(x)$ is the Bessel function of the second kind, and $g_{X}$ is the number of internal degrees of freedom of particle $X$.
Specifically, $g_{H,A,H^+,H^-,\chi}=1$.
 The decay rates that enter into $\Gamma\left(X\to N_{1} \, \ell\right)$ are calculated as 

\begin{align}\label{decaytoN1}
\Gamma(H\to N_1 \bar{\nu}_\alpha )&=\frac{(m_H^2-m_{N_1}^2)^2 }{32\pi m_H^3}(f^2_{1\alpha}\cos^2\theta+2h^2_{1\alpha}\sin^2\theta),\\
\Gamma(\chi\to N_1 \bar{\nu}_\alpha )&=\frac{(m_\chi^2-m_{N_1}^2)^2 }{32\pi m_\chi^3}(2h^2_{1\alpha}\cos^2\theta+f^2_{1\alpha}\sin^2\theta),\\
\Gamma( A\to N_1\,\bar{\nu}_\alpha)&=\frac{(m_A^2-m_{N_1}^2)^2  }{32\pi m_A^3}
f^2_{1\alpha},\\ 
\Gamma( H^+\to N_1\,\bar{\ell}_\alpha)&=\frac{(m_{H^+}^2-m_{N_1}^2)^2 }{32\pi m_{H^+}^3}f^2_{1\alpha}.
\end{align}

\begin{figure}[!htbp]
\begin{centering}
\includegraphics[width=.4\textwidth]{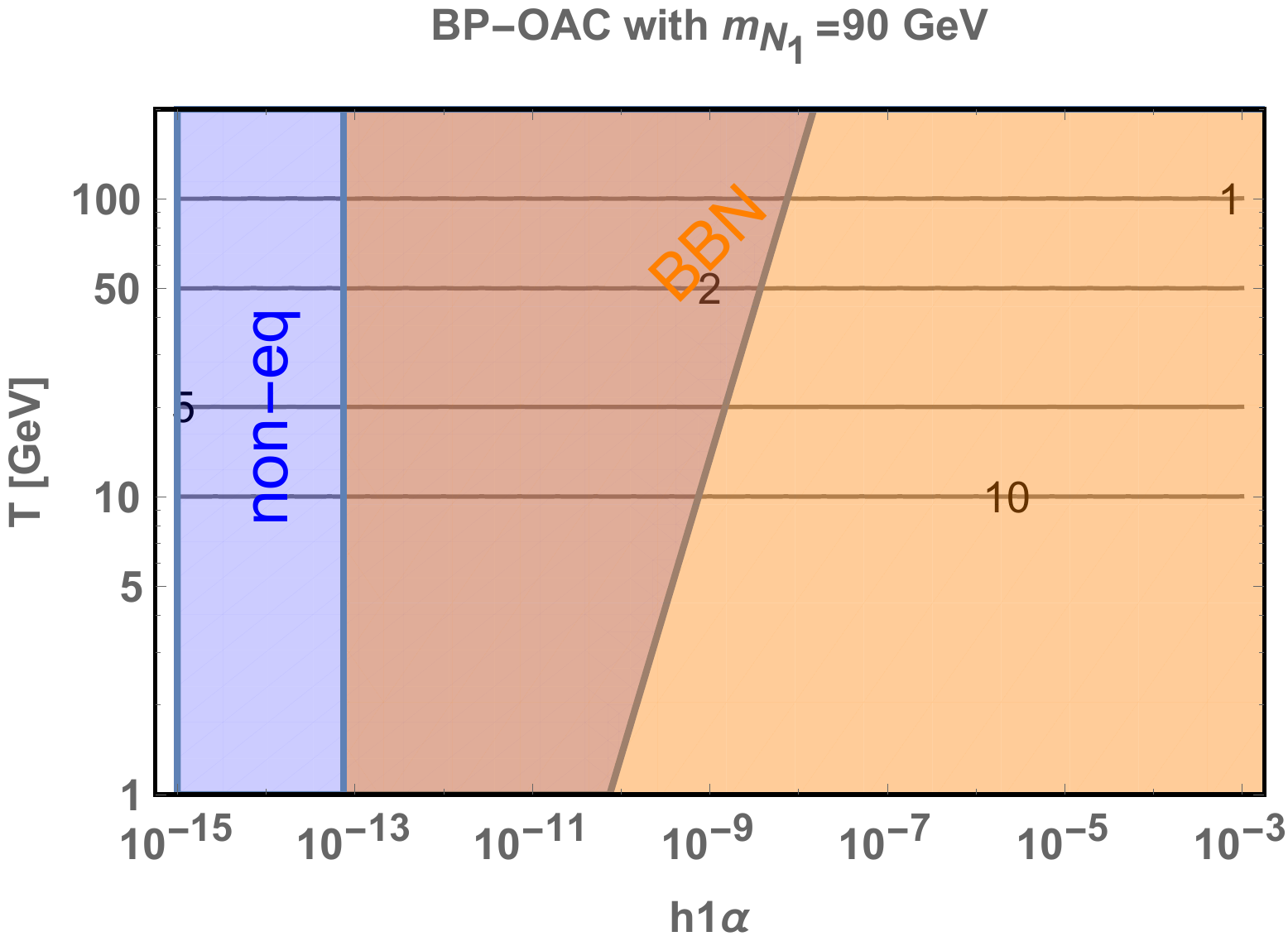}
\includegraphics[width=.42\textwidth]{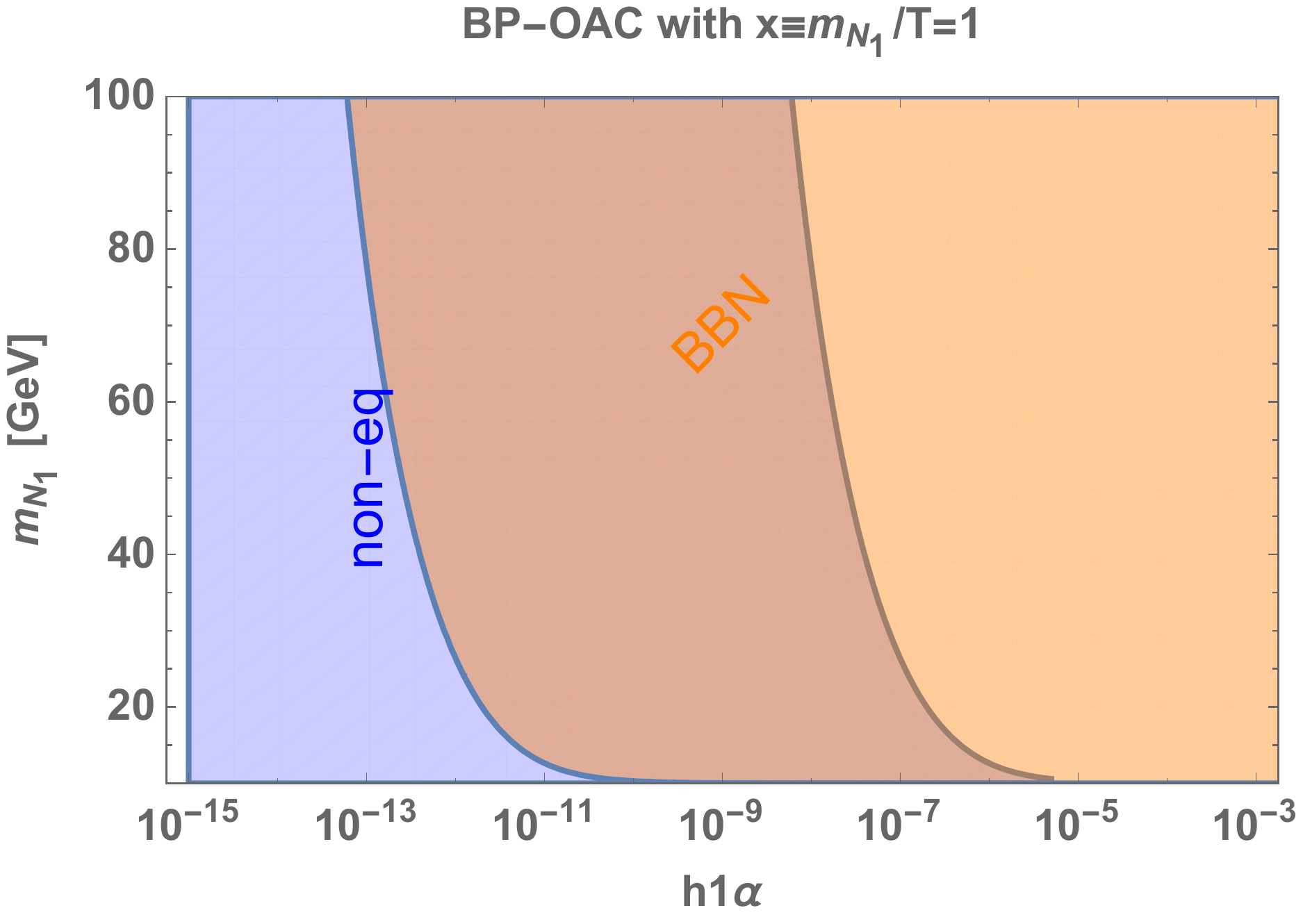}
\caption{The decouple condition bounds on the $h_{1\alpha}$ for the pattern of $O\to A\to C$ corresponds to Fig.~\ref{fig:GWsing1}. The $z=m_{N_1}/T$ is shown by black lines.} \label{fig:dec}
\end{centering}
\end{figure}

Firstly, the requirements that the $N_1$ does not reach thermal equilibrium, i.e., $\Gamma(X\to N_1 \nu)<H(T)$, set the upper limits on the Yukawa couplings $h_{1\alpha}$, and the lower limits on which is set by the Big-Bang nucleosynthesis (BBN) requirements~\cite{bbn}. At the BBN epoch, the finite temperature effects are negligible. 
 We note that the increase of DM mass $m_{N_1}$ means a larger parameter spaces allowed by the non-equibrium conditions and the BBN, as shown in Fig.~\ref{fig:dec}. 
As for the Yukawa coupling of $f_1$, one needs to worry about the disturbation to the BBN or CMB by the decay of dark scalars to $N_1$, because all these dark scalars are in equilibrium before $T_{fo}\sim m_{A,H,H^\pm}/26$ and these particles' number density quickly drops to be almost zero after $T_{fo}$.
 
\begin{figure}[!htbp]
	\begin{centering}
	\includegraphics[width=.41\textwidth]{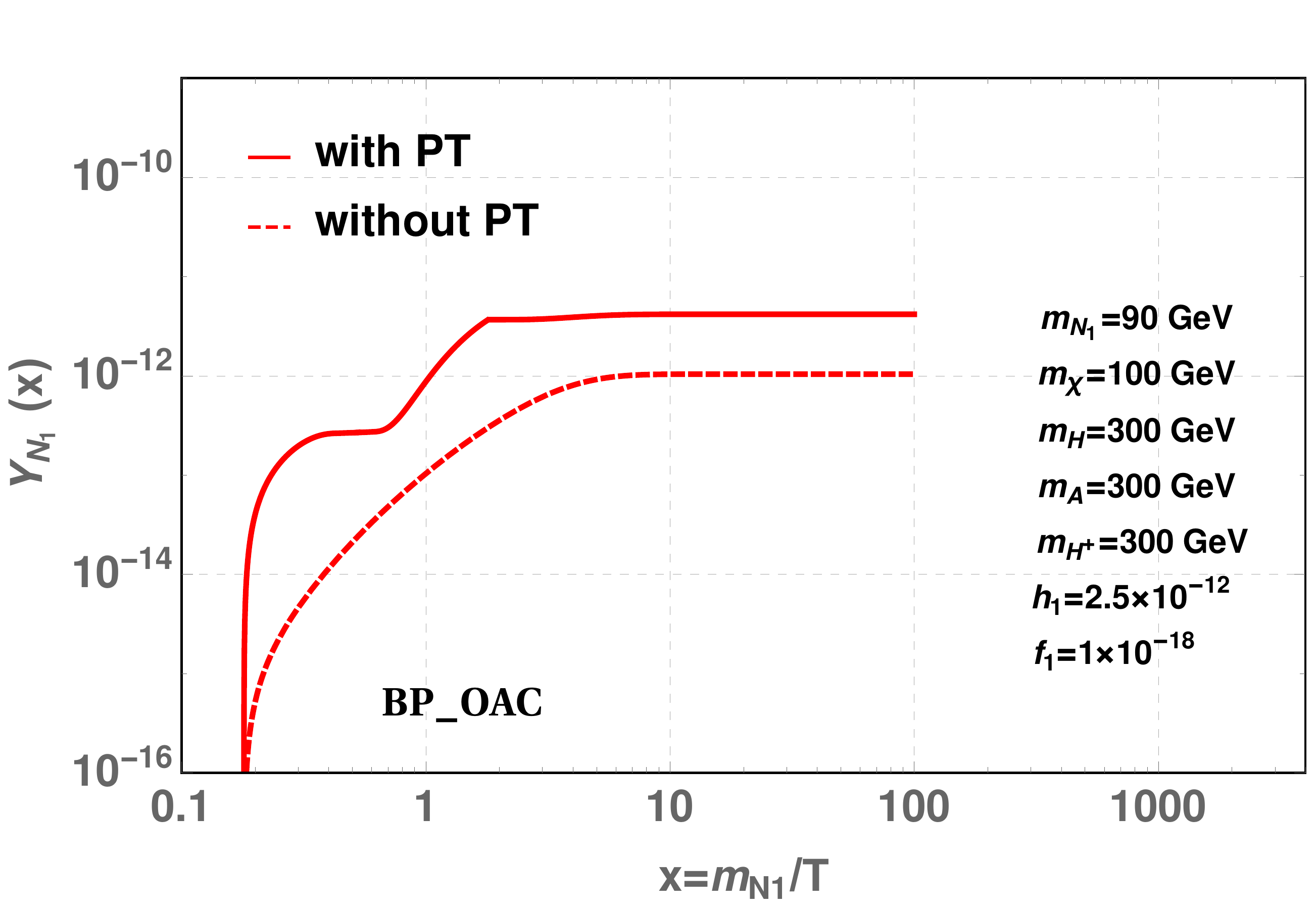}
	\includegraphics[width=.42\textwidth]{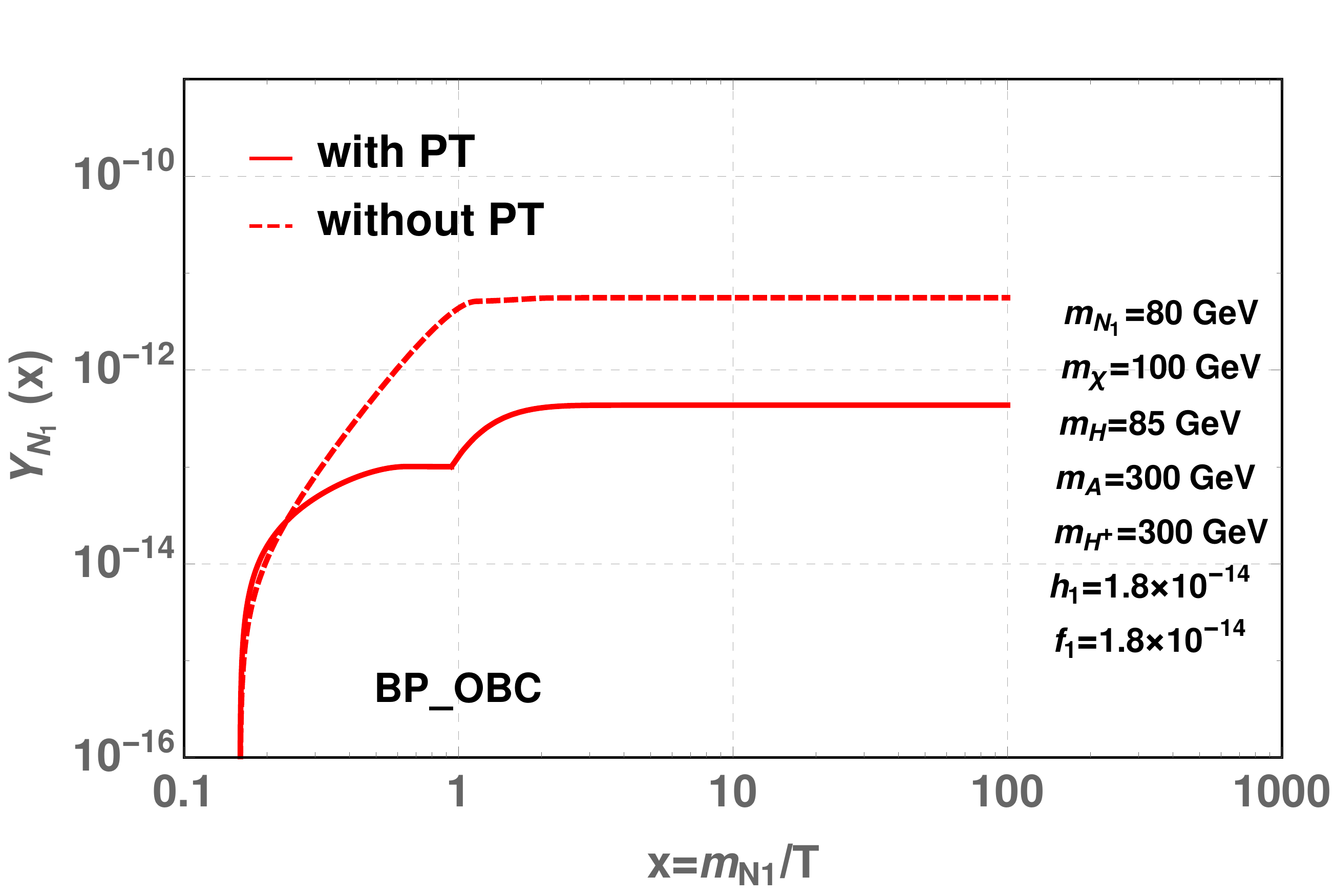}
	\caption{Relic density of the dark matter $N_1$ in the benchmark model of ${\rm O\to A\to C}$ (left panel) and ${\rm O\to B\to C}$(right panel) with and without thermal effects, $f_1$ and $h_1$ represent the yukawa couplings of the first generation of $N_1$.} \label{fig:N_bp5c2}
	\end{centering}
\end{figure}

Different from the $O\to A\to C$ benchmark,
in the benchmark $O\to B\to C$, we have the contribution of the $2\leftrightarrow 2$ process $\gamma H^\pm \leftrightarrow \ell N_1$ dominates over the decay/inverse decay process $H/A/H^\pm \leftrightarrow N\ell$ when one does not include the thermal mass correction.
During the epoch between $T_H$ (the first-stage second-order phase transition of $Z_2$ phase) and $T_n$ (the second-stage first-order phase transition of EW phase), 
we have a null value for $m_{H^\pm}(T)$, where the contribution of the $H/A\to N \ell $ process can dominate over the $2\leftrightarrow 2$ process contribution to the DM production depending on if the kinematical threshold is allowed. Before the $Z_2$ phase is broken($T> T_H$), i.e., in the symmetric phase, and after the EW phase is broken ($T<T_n$), the  $2\leftrightarrow 2$ process can dominate over the $H/A\to N \ell $ process once the kinematical threshold is allowed. 
The Boltzman equation for this scenario is
\begin{eqnarray}
\frac{dY_{N}}{dx}\hspace{-0.1cm}&=&\frac{1}{sxH}\Bigg[\left(\frac{g_X m_X^2m_N\Gamma_{X\to N\nu/\ell}}{2\pi^2x}\right)K_1\left(\frac{m_X}{m_N}x\right) \nn\\
&&+\frac{g_{\gamma}g_{H^\pm}}{32 \pi^4}\frac{m_N}{x}\int_{m_{H^\pm}^2}^\infty d s  4 p_{\gamma H^\pm}^2 \sigma _{\gamma H^\pm\to \ell N}K_1 (\frac{x\sqrt{s}}{m_N})\Bigg] \;,
\label{eq:bltmYchiN1OBC}
\end{eqnarray}
here the $p_{\gamma H^\pm}=(s-m_{H^\pm}^2)/2\sqrt{s}$, and the cross section of $\sigma _{\gamma H^\pm\to \ell N}$ is given in Appendix~\ref{app:xec}.

In Fig.~\ref{fig:N_bp5c2}, we show the entropy normalized number density $Y_{N_1}$ as a function of temperature for fixed $m_{N_1}$. There, the mixing angles effects are negligible. In the $O\to A\to C$ and $O\to B\to C$  scenarios all dark scalars should live in thermal bath to avoid disturbing the BBN. In the $O\to B\to C$ scenario, the dominant channel is the scattering process and therefore the results do not rely on the mixing angle. 
  The curve ``with(out) PT" stands for the results with(out) taking into account the thermal effects. In this situation, all the dark scalars live in the thermal bath before the freeze out due to the sizable DM-SM couplings. The left panel of Fig.~\ref{fig:N_bp5c2} ($O\to A\to C$) is obtained by solving Eq.~\ref{eq:decayDM} with the thermal masses as function of temperature shown in Fig.~\ref{fig:thmOAC}. The right panel of  Fig.~\ref{fig:N_bp5c2} ($O\to B\to C$) is obtained by solving the Eq.~\ref{eq:bltmYchiN1OBC} (all the thermal masses adopted here are the same as Fig.~\ref{fig:thmOBC}). 
The departure of the ``with PT" yield $Y_{N_1}(x)$ from the ``without PT" yield is noticeable, this is because the kinematical thresholds for the decay and scattering process are dynamical during the temperature evolution.
 In particular, the thermal mass of $A,H^\pm$ mostly smaller than the corresponding zero temperature masses within the temperature epoch under study. The two plateau behaviors of the solid curves reflect if the thermal masses reach the threshold of the DM production processes during the two stage phase transition.

\subsection{Freeze-in $N_1$ augmented by late decay of FIMP $\chi$}
\label{sec:N1coa}

In this section, we study the case that DM $N_1$ is partially generated by freeze-in mechanism, and partially generated by
late decay of FIMP $\chi$, in the phase transition pattern $O\to B\to C$. 
When the yukawa 
couplings $f_{1\alpha}\sim h_{1\alpha}\ll \mathcal{O}(1)$ and $\lambda_{s\phi,s\eta},\mu_{soft}\ll 1$, 
both $\chi$(in this case mostly $S$ component) and $N_1$ never reach thermal equilibrium(both are FIMPs). 
Then the coupled Boltzmann equations are
\begin{eqnarray}
\frac{dY_\chi}{dx}\hspace{-0.1cm}&=&\hspace{-0.1cm}\frac{1}{sxH} \left[\left(\frac{g_{H^\pm(A)} m_{H^\pm(A)}^2m_\chi   }{2\pi^2x}\right) \Gamma_{H^\pm(A)\to \chi W^\mp (Z) }K_1\left(\frac{m_{H^\pm(A)}}{m_\chi}x\right)  -s \Gamma(\chi \to N_1\nu_\alpha ) Y_\chi \right]\;,
\nn\\
\frac{dY_{N}}{dx}\hspace{-0.1cm}&=&\frac{1}{sxH}\Bigg[\left(\frac{g_{H^\pm,A,H} m_{H^\pm,A,H}^2m_N\Gamma_{H^\pm,A,H\to N\ell}}{2\pi^2x}\right)K_1\left(\frac{m_{H^\pm,A,H}}{m_N}x\right)+s\Gamma(\chi \to N_1\nu_\alpha )  Y_\chi\nn\\
&&+\frac{g_{\gamma}g_{H^\pm}}{32 \pi^4}\frac{m_N}{x}\int_{m_{H^\pm}^2}^\infty d s  4 p_{\gamma H^\pm}^2 \sigma _{\gamma H^\pm\to \ell N}K_1 (\frac{x\sqrt{s}}{m_N})\Bigg]
 \;,
\label{eq:bltmYchiN1}
\end{eqnarray}
The decay widths are given as follows, 
\begin{eqnarray}
&\Gamma(H^\pm\rightarrow W^\pm\chi)=\frac{e^2(m_{H^+}^4+(m_\chi^2- m_W^2)^2- 2m_{H^+}^2 (m_\chi^2+m_W^2) )^{3/2} \sin^2\theta}{64 \pi  m_W^2 \sin^2\theta_w m_{H^+}^3}\; ,\label{eq:ind2}\\
&\Gamma(A\rightarrow Z\chi)=\frac{e^2(m_{A}^4+(m_\chi^2- m_Z^2)^2- 2m_{A}^2 (m_\chi^2+m_Z^2) )^{3/2}\sin^2\theta}{64 \pi  m_Z^2 \cos^2\theta\sin^2\theta_w m_{A}^3}\; ,\label{eq:ind1}\\
&\Gamma(\chi \to N_1\nu_\alpha )=\frac{(m_{\chi}^2-m_{N_1}^2)^2}{64\pi m_{\chi}^3}
\Big( 2 h^2_{1 \alpha} \cos^2\theta + f^2_{1 \alpha}\sin^2\theta \Big) \;.
\end{eqnarray}

\begin{figure}[!htbp]
	\begin{centering}
	\includegraphics[width=.4\textwidth]{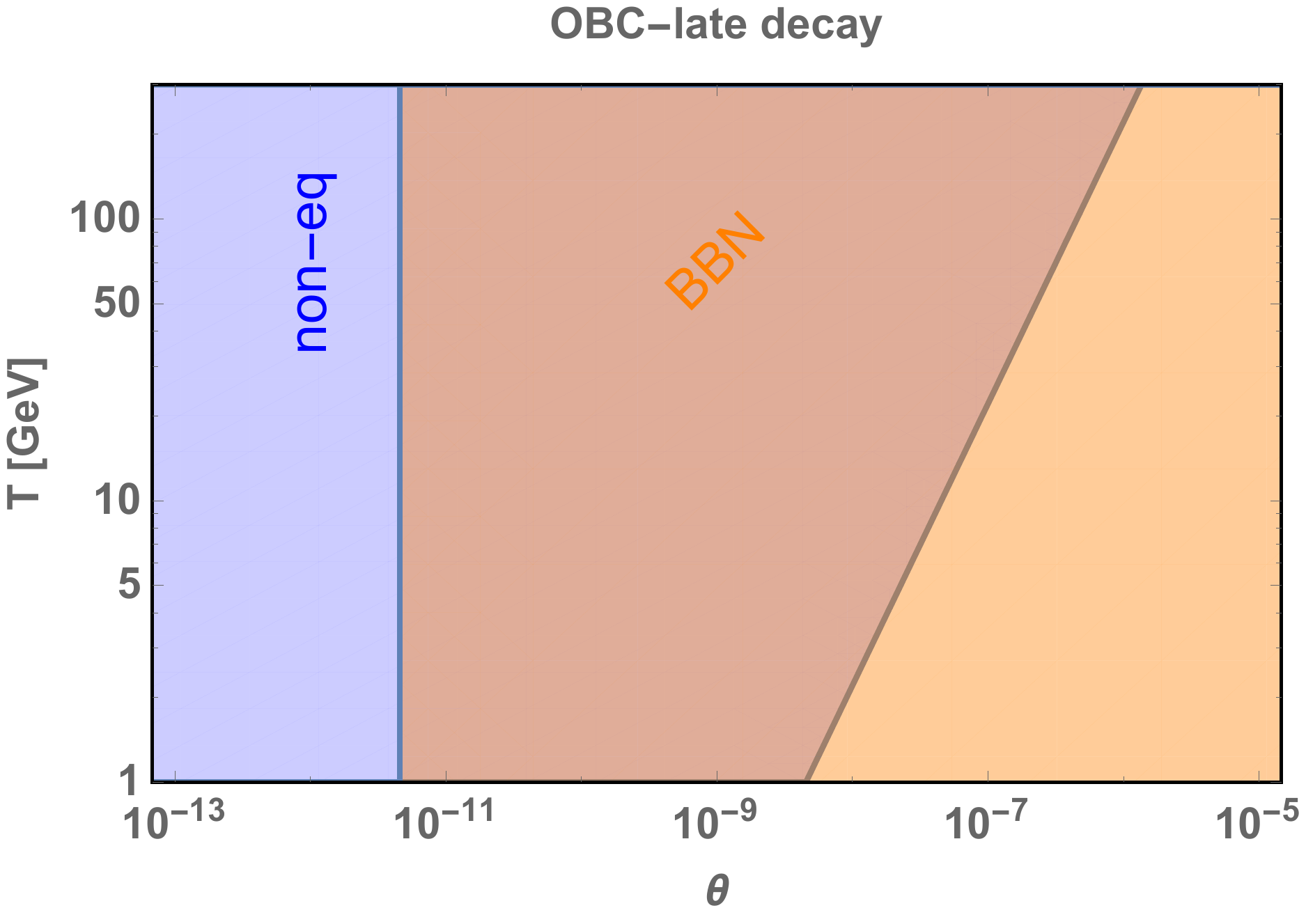}
		\includegraphics[width=.4\textwidth]{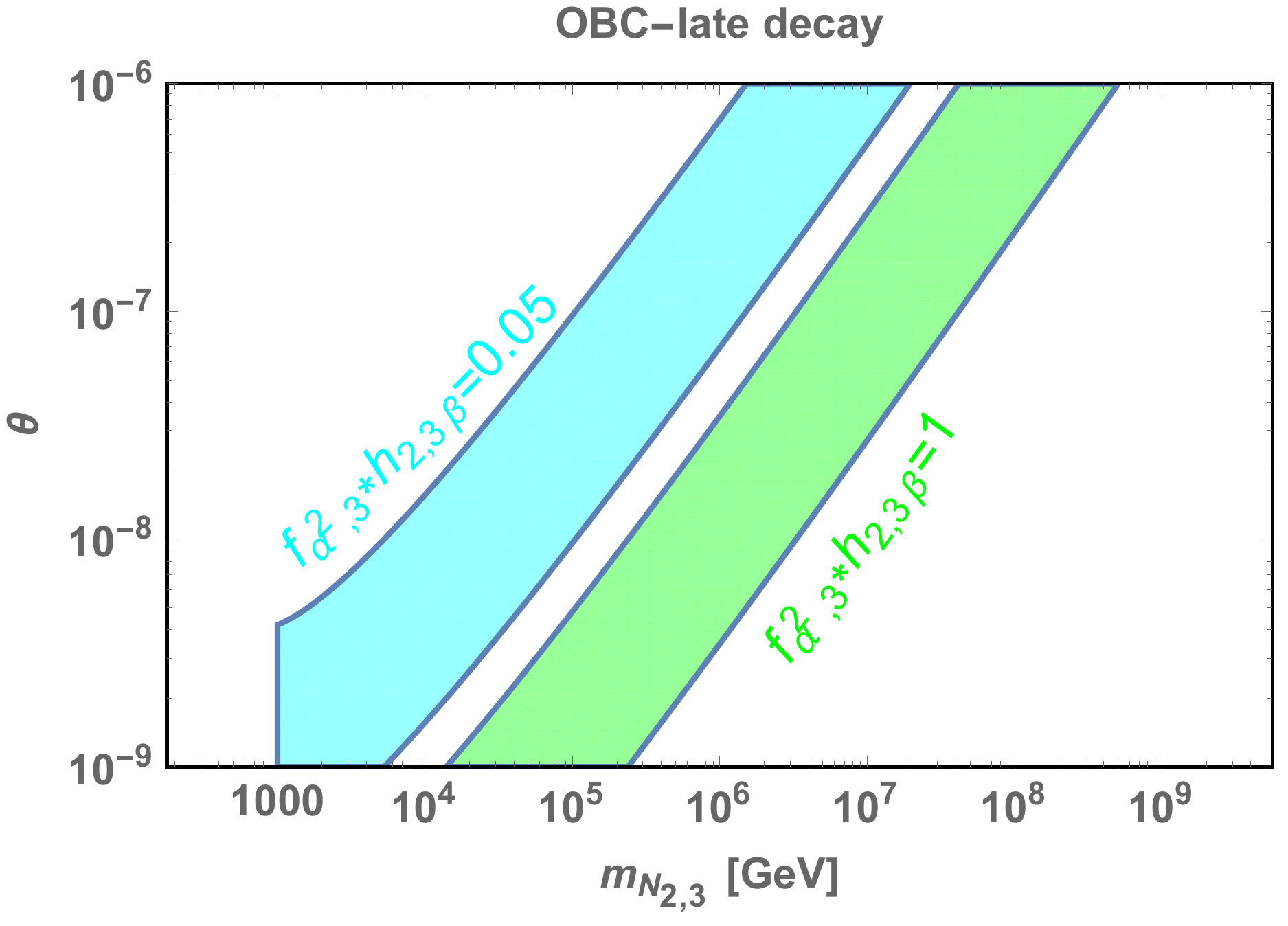}
	\caption{Left: Parameter spaces of $\theta$ allowed by non-equilibrium conditions and BBN of the late decay of FIMP $\chi$ dominated DM production in scenario of ${\rm O\to B\to C}$. Right: The neutrino mass allowed parameter spaces of $\theta$ and $m_{N_{2,3}}$ from Eq.~\ref{eq:neumM}, here we set the range of $10^{-2}-10^{-1}$ eV for the neutrino mass.} \label{fig:N_bpOBC}
	\end{centering}
\end{figure}

The $\chi$ particle mostly comes from the inert sectors decay as given by Eqs.~(\ref{eq:ind2},\ref{eq:ind1}), the decouple conditions is given by $\Gamma( H^\pm/A\to\chi W^\pm/Z)<H(T)$. The left panel of Fig.~\ref{fig:N_bpOBC} depicts that 
the combination of non-equilibrium condition and BBN bounds the mixing angle $10^{-11}<\theta<10^{-6}$ depending on the temperature. 
To obtain the correct magnitude of the neutrino masses, as aforementioned in Sec.~\ref{S:SIDM}, the mixing angle can not be infinitely small. 
For the $\theta$ allowed by the non-equilibrium conditions and the BBN, the Yukawa couplings of $f_{2,3}$ and $h_{2,3}$ and the heavy neutrino masses are restricted, as indicated by Eq.~\ref{eq:neumM}.
So we illustrate the constraints in the right panel of Fig.~\ref{fig:N_bpOBC},  the mixing angle $\theta$ and the heavy neutrinos mass are restricted into a narrow regions for typical values of the Yukawa couplings $f_{\alpha 2,3}$ and $h_{2,3\beta}$.

\begin{figure}[!htbp]
	\begin{centering}
		\includegraphics[width=.325\textwidth]{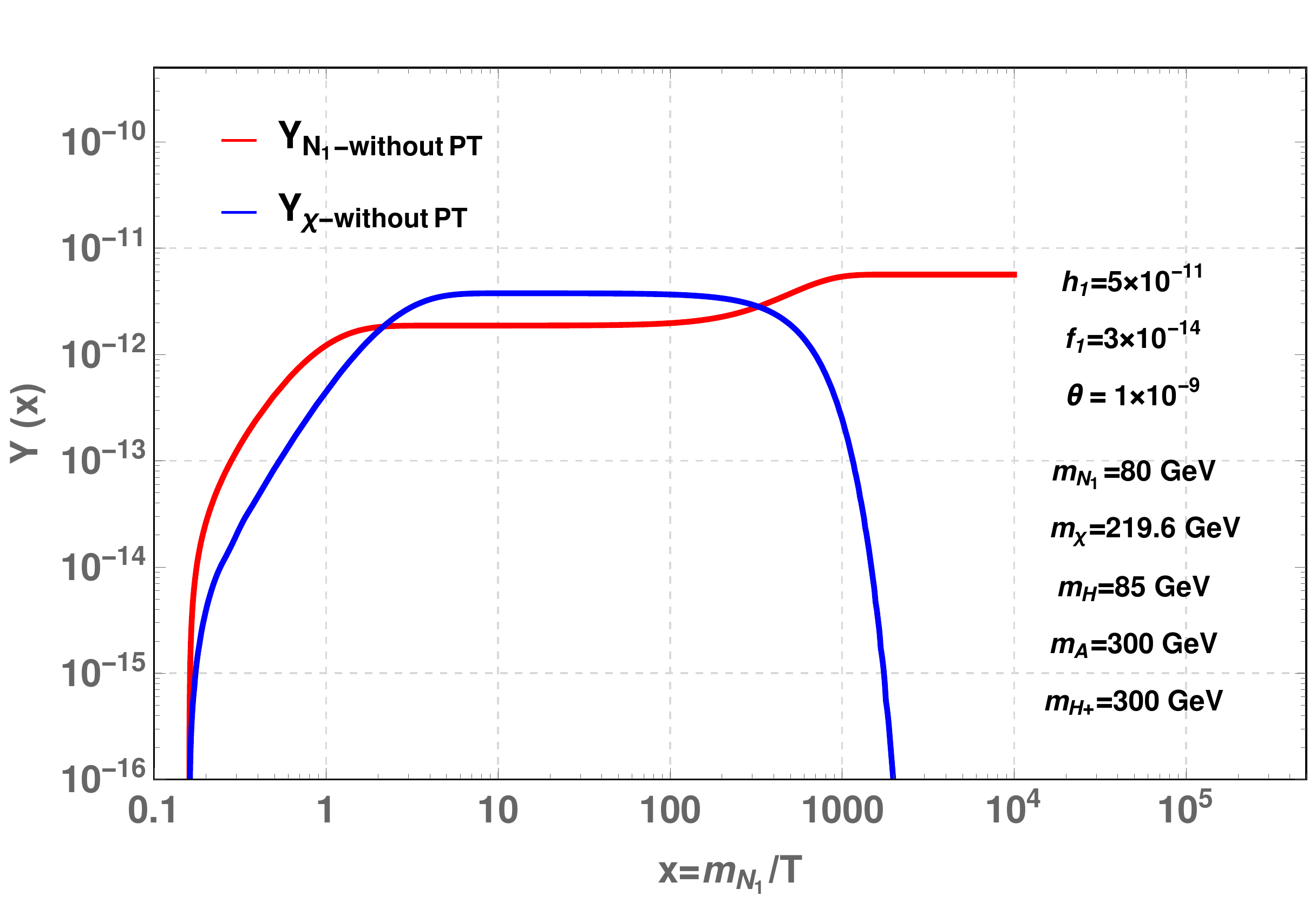}
		\includegraphics[width=.325\textwidth]{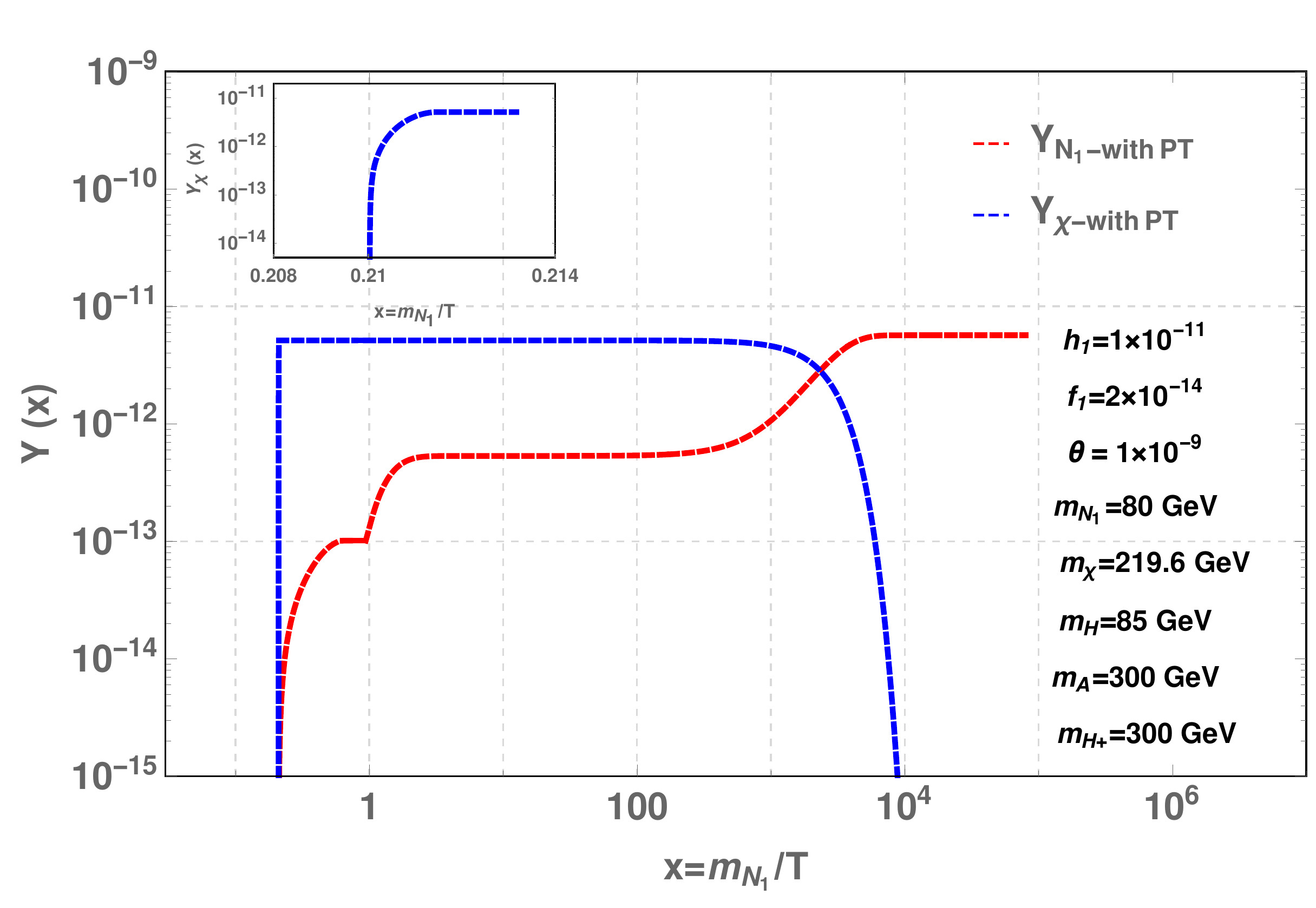}
	\includegraphics[width=.325\textwidth]{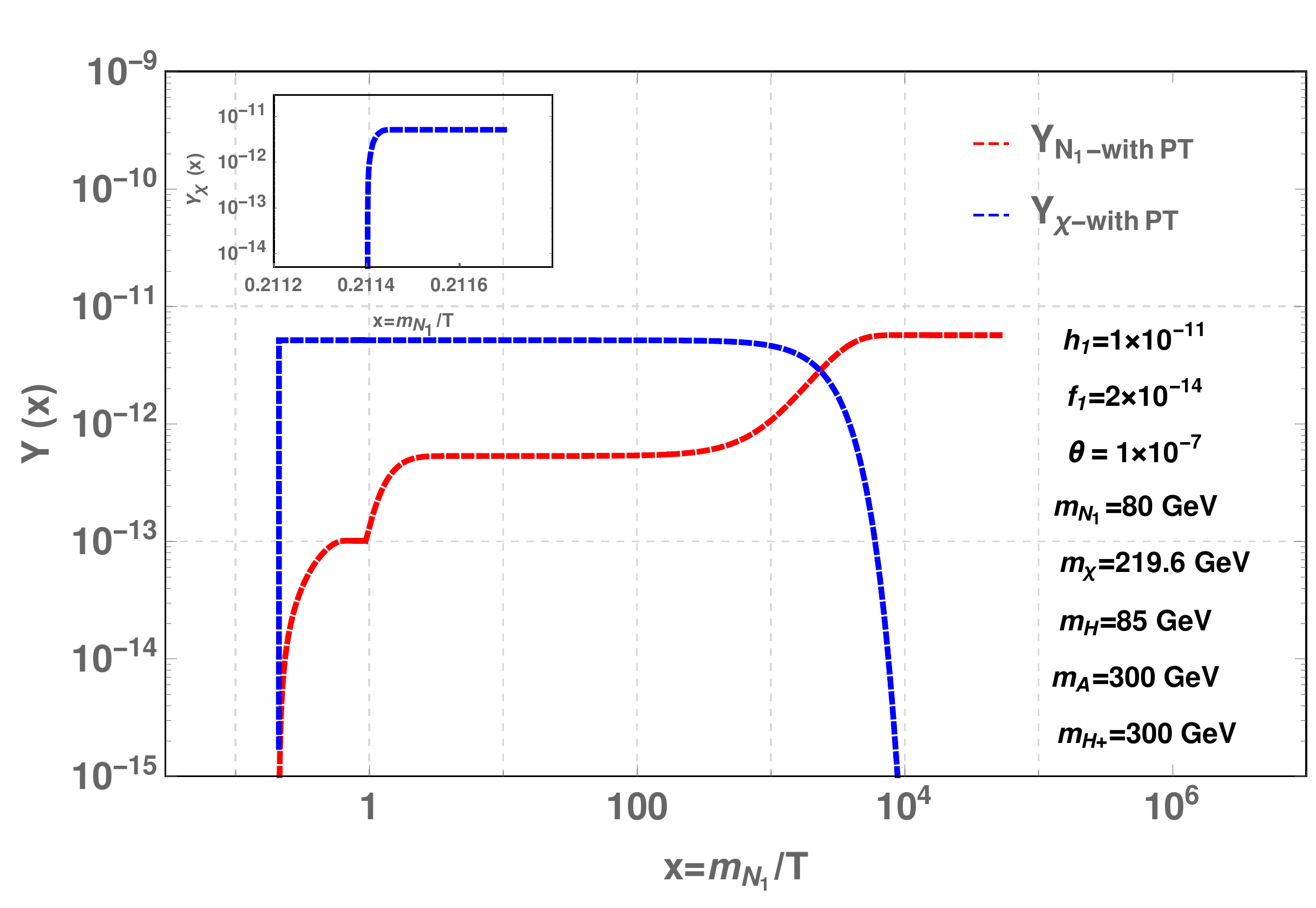}
	\caption{Relic density of the dark matter $N_1$ in the benchmark model of ${\rm O\to B\to C}$ without thermal effects (left) and with thermal effects by taking $\theta=10^{-9}$ (left and middle panels) and $\theta=10^{-7}$ (right panel). The insets illustrate the rapidly growth of $Y_\chi$ at the early stage. Same as Fig.~\ref{fig:N_bp5c2}, the $f_1,h_1$ are the yukawa couplings of the first generation $N_1$.} \label{fig:N_bpOBCth}
	\end{centering}
\end{figure} 

We consider the $\chi$ particle mostly comes from the decay of $H^\pm$, then the late decay $\chi\to N_1\nu$ dominantly contributes to the relic density of $N_1$.  
Before that, the contribution of the annihilation process dominates the production of $N_1$, as depicted in the right panel of Fig.~\ref{fig:N_bp5c2}. In the left panel of Fig.~\ref{fig:N_bpOBCth} we show the ``without PT" case without taking into account the thermal mass, the decay $H^\pm\to \chi W^\pm$ is always active during the production of $\chi$ particle. 
For the case with the thermal effect labeled as ``with PT" in the right two panels of Fig.~\ref{fig:N_bpOBCth}, the thermal masses $m_{H,A,H^\pm}(T)$ are adopted, and the $m_\chi(T)\approx m_\chi$ since here we need to assume negligible small $\lambda_{s\eta,s\phi},\theta$ to ensure $\chi$ does not enter into the equilibrium. The kinematical threshold of $H^\pm\to \chi W^\pm$ can only open at the pretty early stage at high temperature, which leads to a sharp increase of $Y_\chi$ at the beginning. 
For the cases with the thermal effects, the correct DM relic density highly relay on the reheating temperature due to the modification of the kinematical threshold by the thermal masses, we plot the middle and the right panels of Fig.~\ref{fig:N_bpOBCth} to illustrate this situation. In comparison with the case of ``without PT", the late decay of $\chi$ contributions dominate the DM relic density of $N_1$, the much smaller $f_1$ and $h_1$ are chosen here to make sure the correct relic density $\Omega h^2=0.12$~\cite{Aghanim:2018eyx} can be obtained. The magnitude of $h_1$ is relatively larger than $f_1$ to ensure the $\chi$ particle can fully decay before BBN.

%------------------%
\section{Gravitational waves from the SFOEWPT}
\label{sec:GWs}
The gravitational wave signals produced during the phase transitions process can be characterized by two parameters of $\alpha$ and $\beta$ at the phase transition temperature $T_\star$(bubble nucleation temperature). 
The first important parameter $\alpha$ is the latent heat normalized by the radiation energy, given by
$\alpha=\epsilon/\rho_{rad}$ with $\rho_{rad}=\pi^2 g_\star T_*^4/30$. The latent heat $\epsilon$ includes the difference of the vacuum energy between the false and true vacuum and the entropy variation $\Delta s$ (see Appendix~\ref{app:entropy}) at the phase transition temperature, given by
\begin{eqnarray}
\epsilon=-\Delta V-T\Delta s=(-\Delta V+ T \frac{\partial V}{\partial T})|_{T=T_\star}\;.
\end{eqnarray}
The second crucial parameter $\beta$ reflects the duration of the phase transition, and characterize the peak frequency of the GW spectrum. Under the assumptions of adiabatic expansion of the universe, one has
\begin{eqnarray}
\frac{\beta}{H_\star}=T_\star \frac{d}{dT}\left(\frac{S_3}{T}\right)\Big{|}_{T_\star}\;,
\end{eqnarray} 
where $S_3$ is the three dimensional Euclidean action for the critical bubble (nucleating bubble).

With the two parameter at hand, one can calculate the gravitional wave signals produced by the SFOEWPT, which includes three dominant contributions:  
bubble collisions, sound waves and Magnetohydrodynamic 
turbulence (MHD) in the plasma~\cite{Caprini:2015zlo,Cai:2017cbj}. The total energy spectrum is given by
\begin{eqnarray}
  \Omega_{\text{GW}}h^2 \simeq \Omega_{\text{col}} h^2 + \Omega_{\text{sw}} h^2 + \Omega_{\text{turb}} h^2\;.
\end{eqnarray}

The first important contribution is the bubble collision, estimating with the envelop approximation~\cite{Kosowsky:1991ua,Kosowsky:1992rz,Kosowsky:1992vn}, which is given by~\cite{Huber:2008hg}
\begin{eqnarray}
  \Omega_{\text{col}} h^2 = 1.67\times 10^{-5} \left(\frac{H_{\ast}}{\beta}\right)^2
  \left(\frac{\kappa \alpha}{1+\alpha}\right)^2 \left( \frac{100}{g_{\ast}} \right)^{1/3}
  \left( \frac{0.11 v_b^3}{0.42 + v_b^2} \right) 
  \frac{3.8(f/f_{\text{env}})^{2.8}}{1+2.8(f/f_{\text{env}})^{3.8}} \ , 
\end{eqnarray}
with the bubble wall velocity $v_b$ and the efficient factor $\kappa$ being functions of the crucial parameter $\alpha$~\cite{Kamionkowski:1993fg},
\begin{eqnarray}
v_b \simeq \frac{1/\sqrt{3} + \sqrt{\alpha^2+2\alpha/3} }{1+\alpha}\;,  \nonumber \quad \quad
\kappa \simeq  \frac{0.715\alpha + \frac{4}{27} \sqrt{3\alpha/2}}{1+0.715\alpha} \;,
\end{eqnarray}
and the peak frequency located at
\begin{eqnarray}
  f_{\text{env}} = 16.5\times 10^{-6} \left(\frac{f_{\ast}}{H_{\ast}}\right)  \left(\frac{T_{\ast}}{100\text{GeV}}\right)
  \left(\frac{g_{\ast}}{100}\right)^{1/6}
  \text{Hz}\;.
\end{eqnarray}
The other two main contributions are the sound waves and the MHD, which are given by
\begin{eqnarray}
  \Omega_{\text{sw}} h^2 & =& 2.65\times 10^{-6} \left(\frac{H_{\ast}}{\beta}\right)
  \left(\frac{\kappa_v \alpha}{1+\alpha}\right)^{2} \left(\frac{100}{g_{\ast}}\right)^{1/3} v_b 
  \left(\frac{f}{f_{\text{sw}}}\right)^3 \left( \frac{7}{4+3(f/f_{\text{sw}})^2} \right)^{7/2} \;\\
  \Omega_{\text{turb}} h^2  &= &3.35\times 10^{-4} \left(\frac{H_{\ast}}{\beta}\right)
  \left(\frac{\kappa_{\text{turb}} \alpha}{1+\alpha}\right)^{3/2} 
  \left(\frac{100}{g_{\ast}}\right)^{1/3} v_b 
  \frac{(f/f_{\text{turb}})^3}{[1+(f/f_{\text{turb}})]^{11/3} (1+8\pi f/h_{\ast})}
\end{eqnarray}
with $\kappa_v \approx \alpha(0.73+0.083\sqrt{\alpha} + \alpha)^{-1}$  and $\kappa_{\text{turb}} \approx 0.1 \kappa_v$ 
describing the fraction of latent heat transformed into the bulk motion of the fluid for sound waves and MHD. Today's Hubble parameter is
\begin{eqnarray}
h_{\ast}=1.65\times 10^{-2} {\text{mHz}}\frac{T_{\ast}}{100 \text{GeV}}(\frac{g_{\ast}}{100})^{1/6}\;.
\end{eqnarray}
 The peak frequencies of sound waves and MHD locate at
\begin{eqnarray}
 && f_{\text{sw}} = 1.9\times 10^{-5} \frac{1}{v_b} 
  \left(\frac{\beta}{H_{\ast}}\right) \left(\frac{T_{\ast}}{100\text{GeV}}\right) 
  \left(\frac{g_{\ast}}{100}\right)^{1/6} \text{Hz} \;,\\
 &&   f_{\text{turb}} = 2.7\times 10^{-5} \frac{1}{v_b} 
  \left(\frac{\beta}{H_{\ast}}\right) \left(\frac{T_{\ast}}{100\text{GeV}}\right) 
  \left(\frac{g_{\ast}}{100}\right)^{1/6} \text{Hz} \;.
\end{eqnarray}

\begin{figure}[!htbp]
\begin{centering}
\includegraphics[width=.6\textwidth]{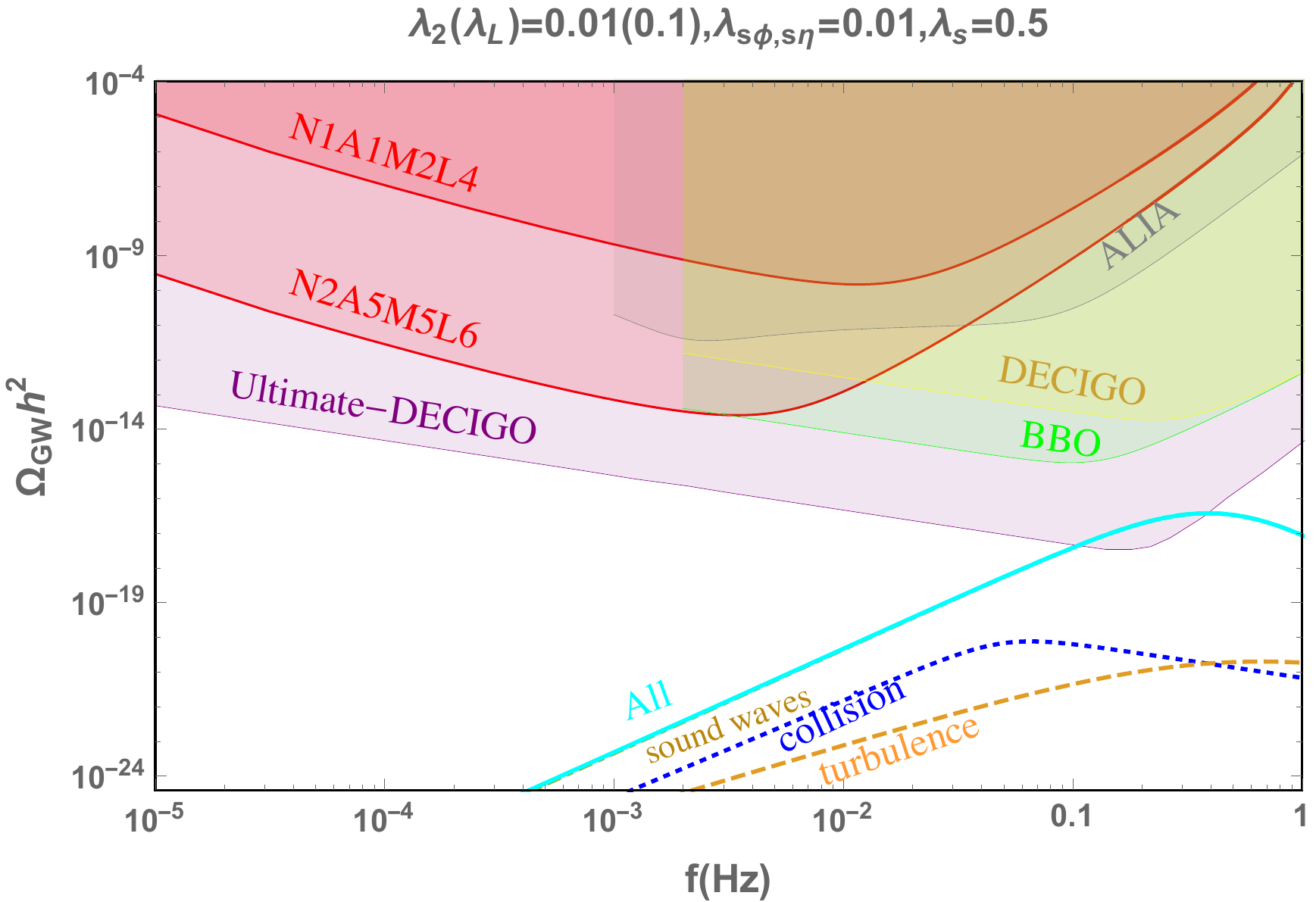}
\caption{The GW signals spectrum for the phase transition pattern of $O\to C$ with $m_{A,H^\pm}=450$ GeV, $m_H(m_\chi)=64(100)$ GeV and the mixing angle is $\theta=0.01$.} \label{fig:GWmix}
\end{centering}
\end{figure}

\begin{figure}[!htbp]
\begin{centering}
\includegraphics[width=.45\textwidth]{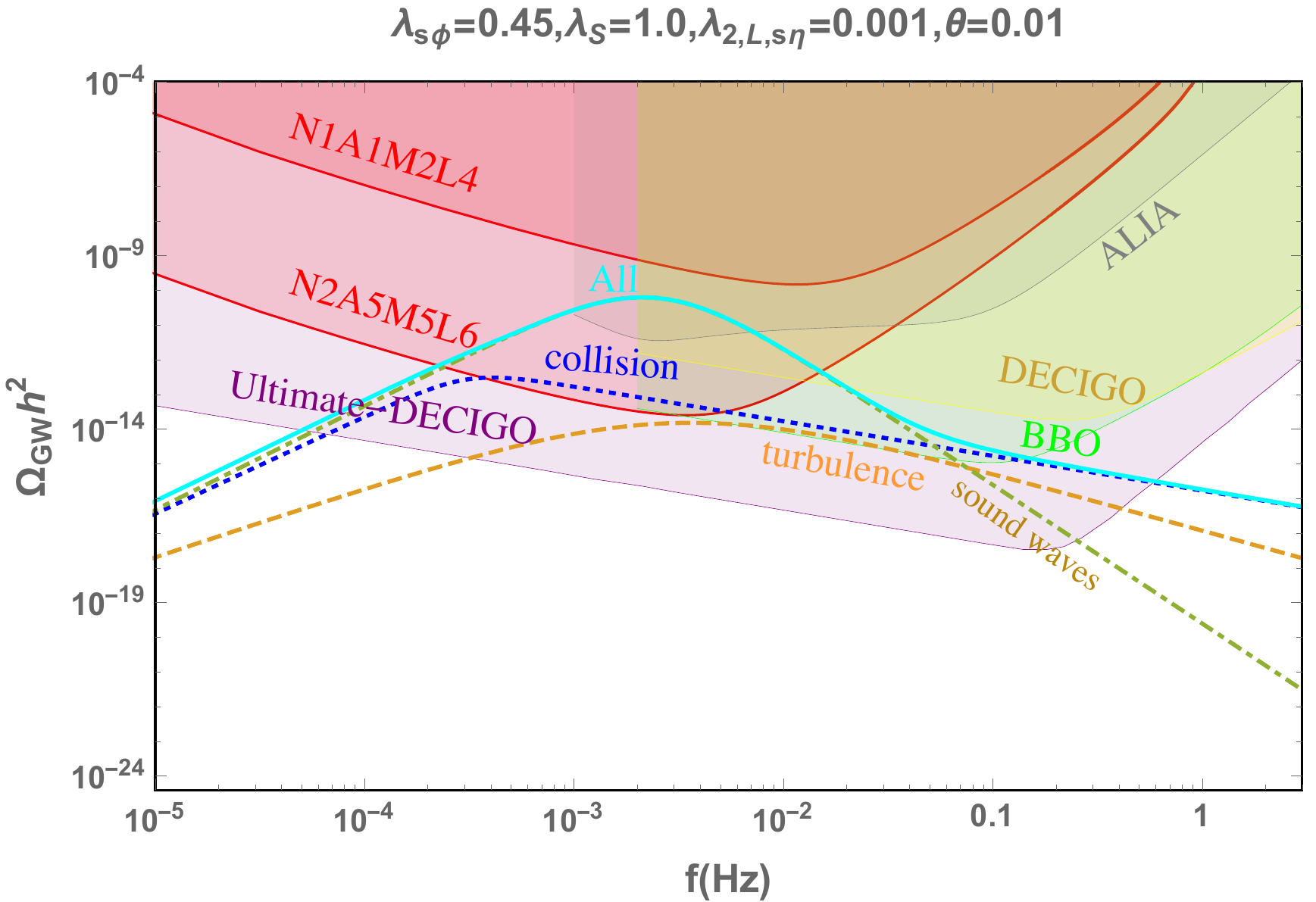}
\includegraphics[width=.45\textwidth]{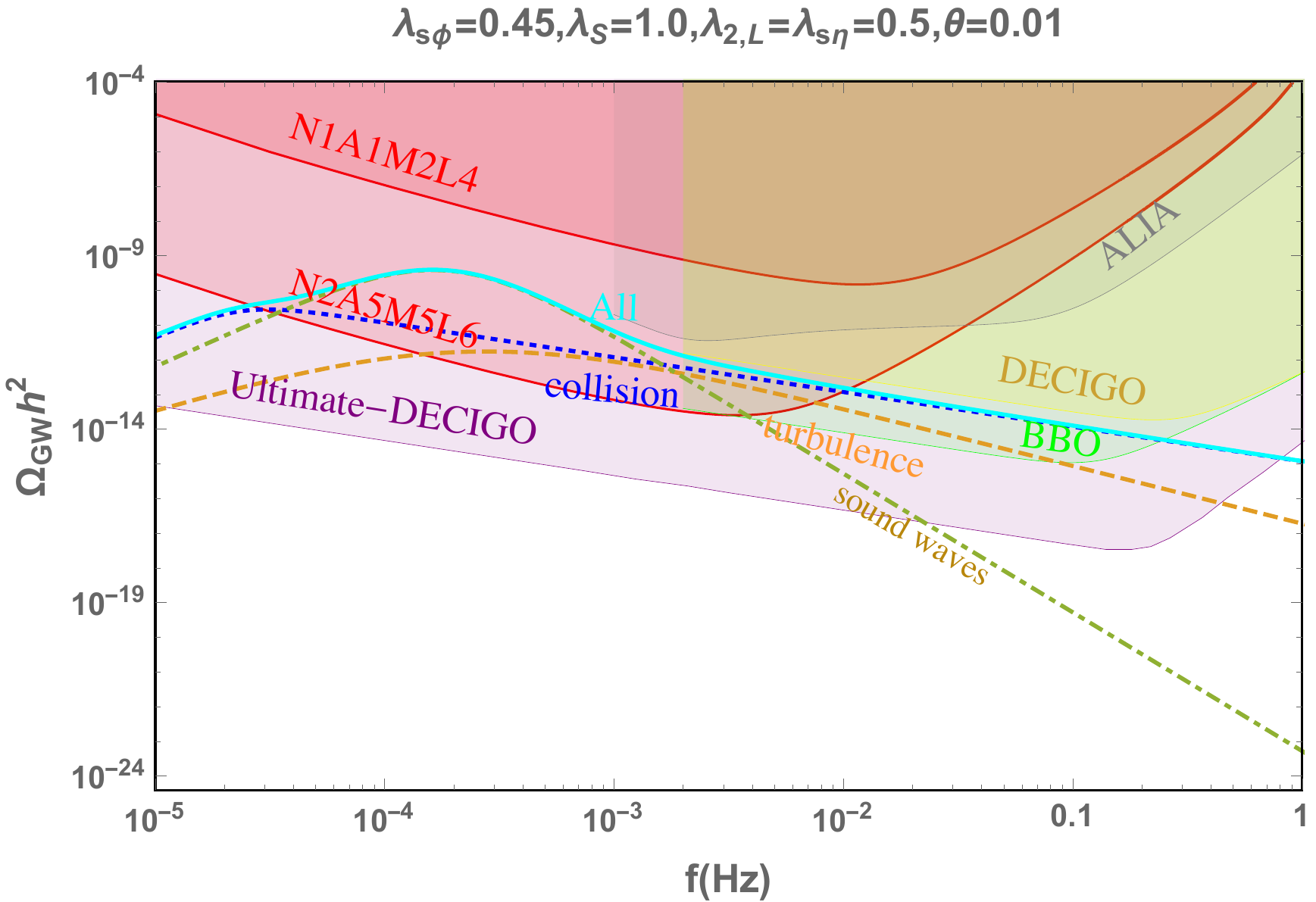}
\includegraphics[width=.45\textwidth]{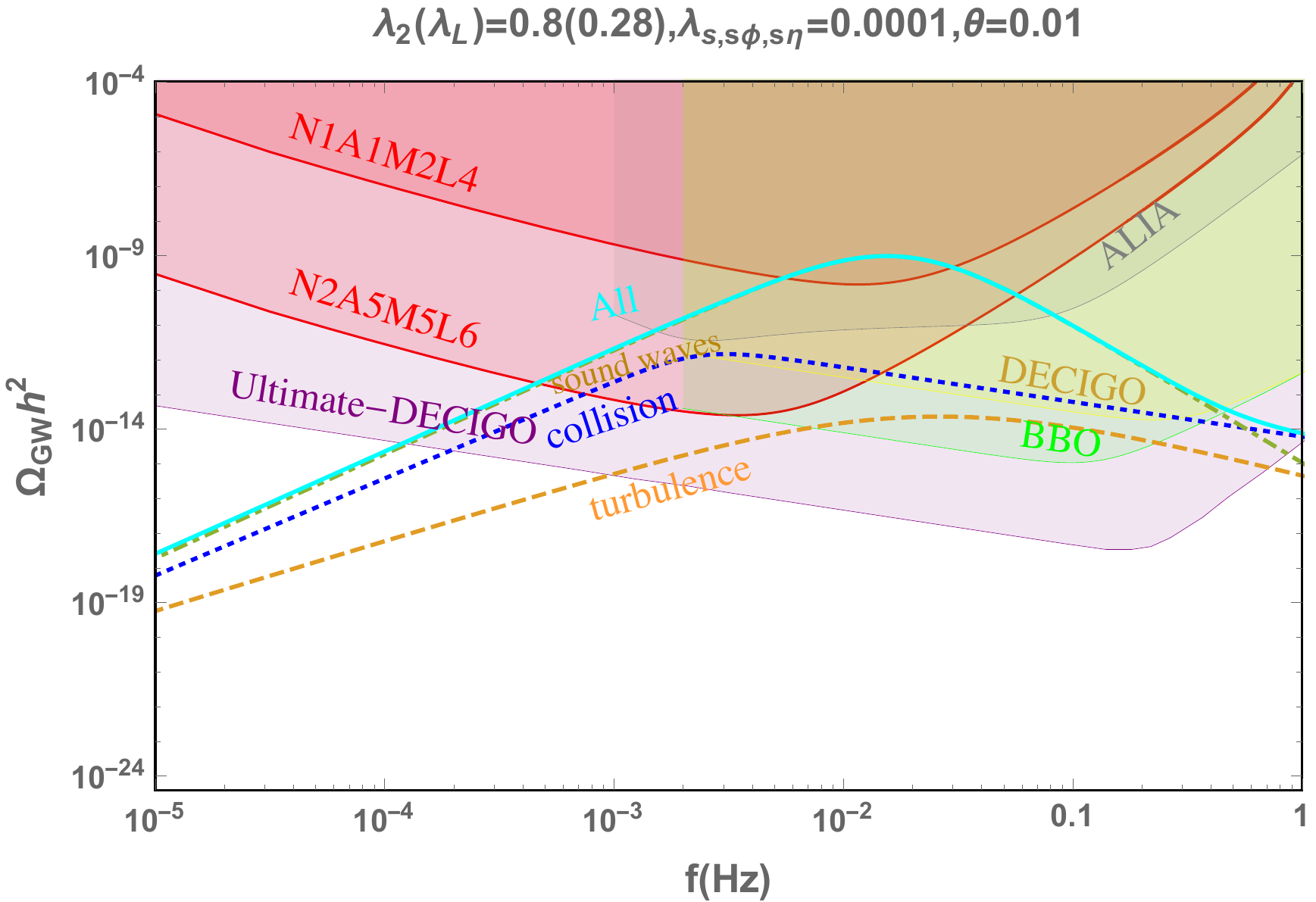}
\includegraphics[width=.45\textwidth]{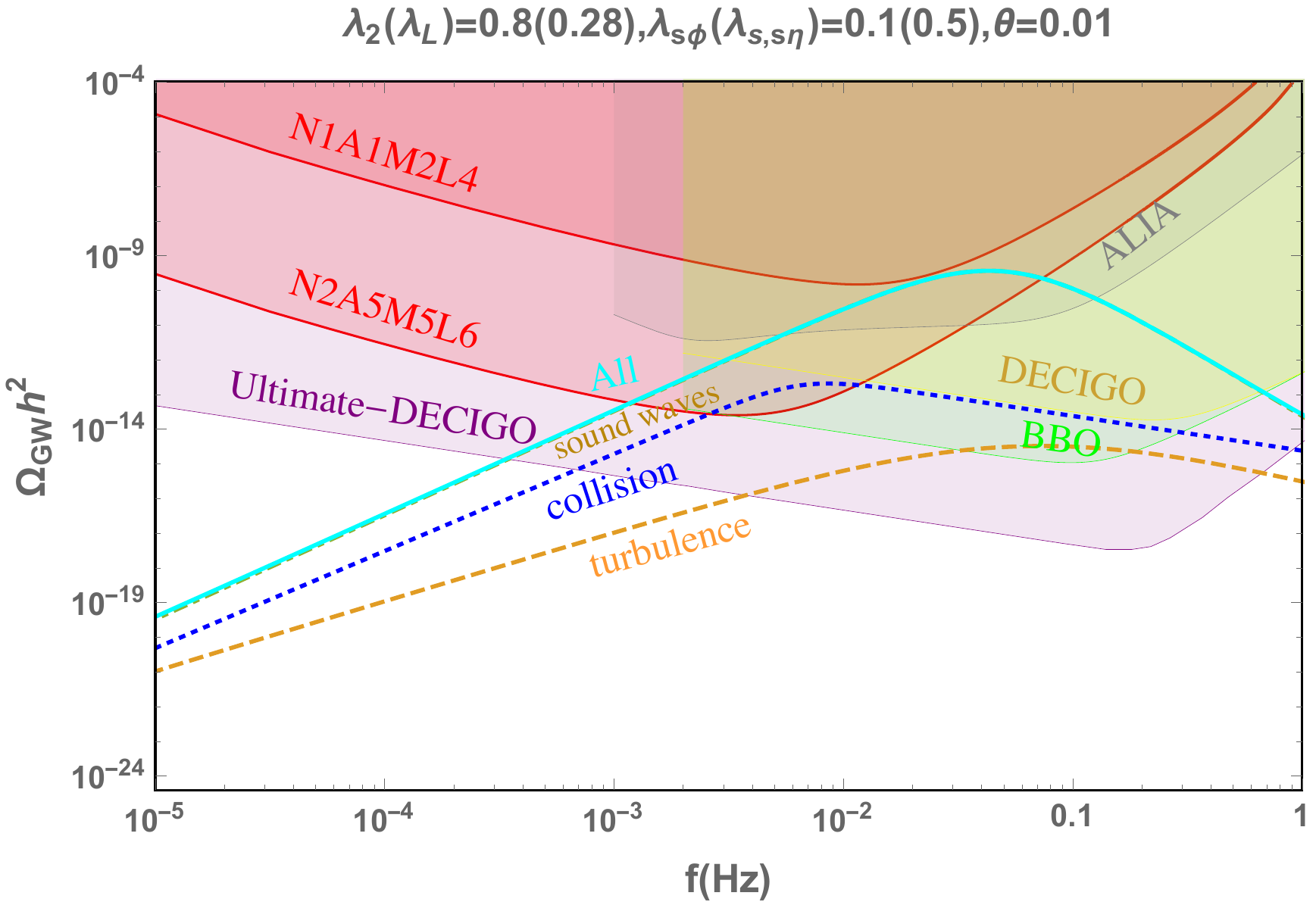}
\caption{The GW signals spectrum for the phase transition patterns of $O\to A\to C$($O\to B\to C$) for up (down) plots with the scalar masses are $m_\chi=100$ GeV and $m_{A,H,H^\pm}=300$ GeV ($m_{A,H^\pm}=300$ GeV and $m_H(m_\chi)= 85 (100)$ GeV).} \label{fig:GWsing1}
\end{centering}
\end{figure}

For the one-step SFOEWPT pattern $O\to C$, one can consider the energy barrier for the phase transition almost coming from the singlet part or the inert doublet part, since the mixing of the two scalars is assumed to be negligible. For the inert doublet dominant case, we refer to Fig.~\ref{fig:GWmix}, which is analogy to the study of Ref.~\cite{Blinov:2015vma}.
While the singlet dominant case requires large Higgs portal coupling $\lambda_{s\phi}$, see Ref.~\cite{Curtin:2014jma}.  
In Fig.~\ref{fig:GWsing1}, we plot the GW signals to be probed by the ALIA, eLISA, BBO, DECIGO and ultimate-DECIGO experiments, which are generated from the two-step SFOEWPT with patterns $O\to A\to C$ and $O \to B \to C$.
The top plots of Fig.~\ref{fig:GWsing1} indicate that the increase of $\lambda_{2,L,s\eta}$ leads to the shift of the frequency to a smaller value. 
The parameter  $\lambda_{2,L,s\eta}$ chosen for the top-right panel ensures that the inert scalars and the singlet scalar live in the thermal bath, which is the benchmark model corresponding to the freeze-in production of $N_1$ with $O\to A\to C$ scenario, 
as studied in left panel of Fig.~\ref{fig:N_bp5c2}. 
The related studies on GWs in this scenario can be also found in Refs.~\cite{Beniwal:2017eik,Kurup:2017dzf,Huang:2018aja,Vaskonen:2016yiu,Ellis:2018mja}\footnote{ For the GW signals from first order phase transitions from a Neutrino mass and Dark Matter model, see Ref.~\cite{Addazi:2017oge,Addazi:2017nmg}.}. 
The bottom plots show that the peak of the spectrum/frequency shifts to the right by increasing $\lambda_{s,s\eta,s\phi}$  for the $O\to B\to C$ pattern. The freeze-in production of $N_1$ without late decay of $\chi$ (the right panel of Fig.~\ref{fig:N_bp5c2}) could be obtained with the parameter setup shown in the bottom-right panel of Fig.~\ref{fig:GWsing1}, except employing a negligible small $\theta$ that will not affect the GW signals production. Considering the freeze-in production of $\chi$, as studied previously in Sec.~\ref{sec:N1coa} (see Fig.~\ref{fig:N_bpOBCth}), the negligibly small $\lambda_{hs,s\phi}$ and $\theta$ are required which also do not affect the phase transition dynamics. This can be realized due to the pattern $O\to B\to C$ is the {\it inert doublet-like} two-step phase transition. The study of the GWs from the inert-doublet two-step phase transition is absent in literature, we fill the blank in this scenario.

\section{Comments on collider searches}
\label{sec:cols}

One of the signals predicted by the SFOEWPT is the deviation 
of the triple Higgs couplings, which might be able to probed at colliders, 
such as the  lepton colliders~\cite{Cao:2017oez,Huang:2015izx} and high energy hadronic colliders~\cite{Arkani-Hamed:2015vfh}. 
We emphasize that this kind of search mostly apply to the SFOEWPT 
through the one-step pattern with a sizable dimensional-six operators after heavy particles being integrated out. 
For the phase transition patterns ${ O\to A\to C}$ and ${O\to B\to C}$, there is no sizable dimensional-six operators could be obtained since there is no contribution from the tree-level. In both cases, the dimensional-six operators can be obtained at one-loop level, see Ref.~\cite{Cheng:2018axr,Carena:2018vpt} for the ${O\to A\to C}$ pattern, and  Ref.~\cite{Blinov:2015vma} for the ${O\to B\to C}$ pattern.  
As argued in Ref.~\cite{Bian:2018mkl}, both the two scenarios might be tested at LHC with the off-shell
$Z$-pair channel~\cite{Goncalves:2017iub}, and at the future lepton colliders~\cite{Cheng:2018axr,Liu:2017gfg,Ellis:2018mja} as well as 100 TeV $pp$ collider~\cite{Curtin:2014jma}. 
As studied in Ref.~\cite{Liu:2017gfg}, the displaced vertex signature might be able to tell the difference between the scalar sectors of this model and the inert doublet model.
We checked the charged Higgs contribution to the Higgs diphoton signal following Ref.~\cite{Liu:2017gfg}, the deviation from the SM prediction is negligible for the FIMP benchmark scenarios under study.

\begin{figure}[!htbp]
\begin{centering}
\includegraphics[width=.5\textwidth]{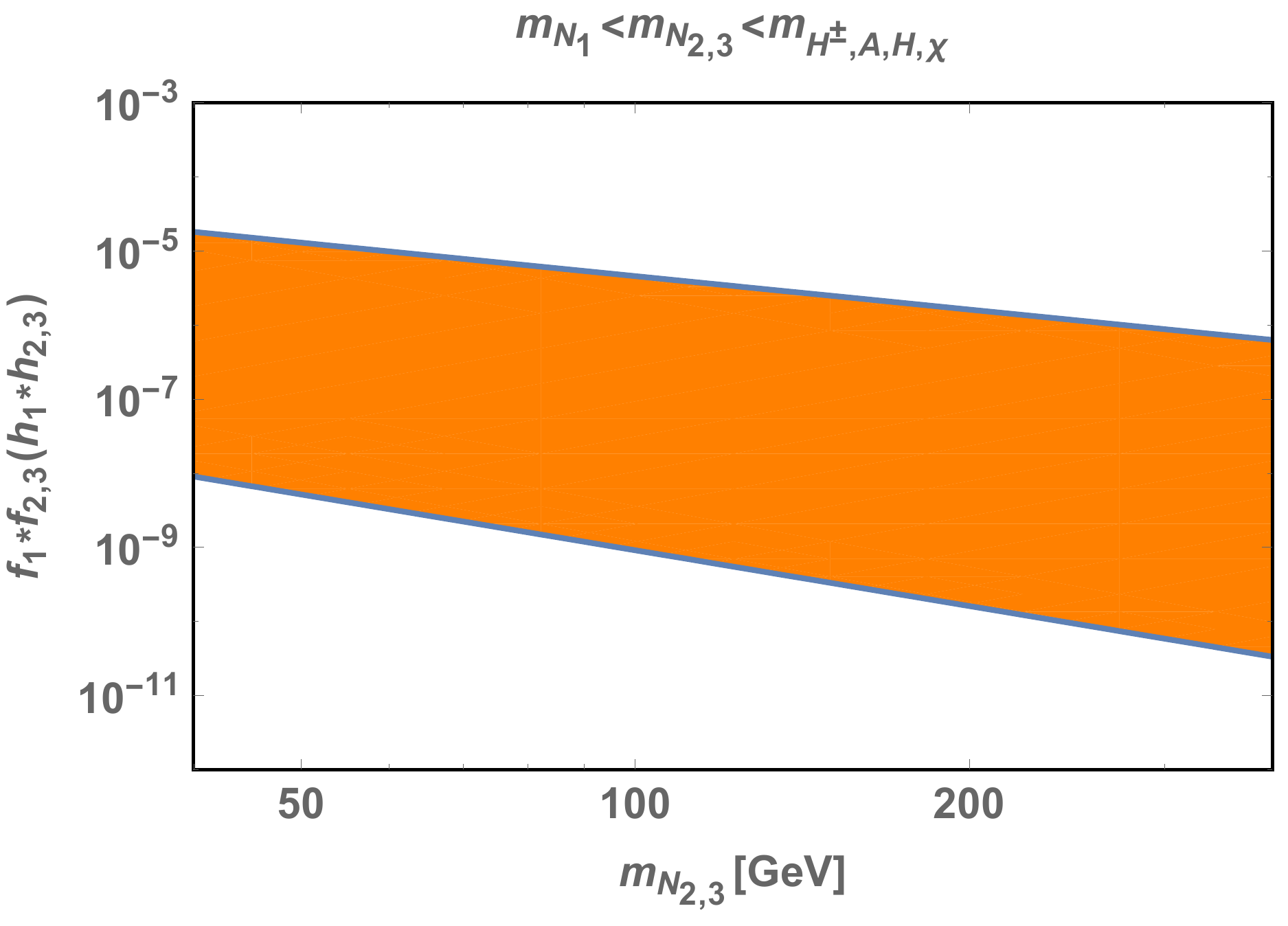}
\caption{BBN allowed parameter spaces of $m_{N_{2,3}}$ and $f_1^2 f_{2,3}^2(h_1^2 h_{2,3}^2)$.} \label{fig:bbn23}
\end{centering}
\end{figure}

For the FIMP DM $N_1$, the decay-lengths of $\chi,H,A,H^\pm$ particles are far beyond the ability of LHC.
For the scenario of $m_{N_{3}}>m_{H^\pm,A,H,\chi}>m_{N_{2}}>m_{N_1}$, as studied in Ref.~\cite{Molinaro:2014lfa}, one can expect the relic abundance of $N_1$ DM coming from the late decay of $N_2$ after the freeze out and before BBN epoch, 
the decays of the heavier singlet fermions $N_{2}\to N_1\,\bar{\ell}\,\ell$ mediated by the scalars (ignore the mixing of $H$ and $\chi$) should happen with $H(T=m_{N_2}/20)>\Gamma_1(N_{2} \to N_{1}  \,\overline{\ell_{\alpha}}\, \ell_{\beta}) > H(T=1$ MeV) and
\begin{eqnarray}
	\Gamma_1(N_{2,3} \to N_{1}  \,\overline{\ell_{\alpha}}\, \ell_{\beta}) 
	\,&=&\, \frac{m_{N_{2,3}}^{5}}{6144\,\pi^{3}\,m_{H^\pm,H,A}^{4}}\left(\left|f_{\alpha 1} \right|^{2} \left|f_{\beta 2,3} \right|^{2}+\left|f_{\beta 1} \right|^{2} \left|f_{\alpha 2,3} \right|^{2}\right)\,,\\
\Gamma_2(N_{2,3} \to N_{1}  \,\overline{\ell_{\alpha}}\, \ell_{\beta}) 
	\,&=&\, \frac{m_{N_{2,3}}^{5}}{6144\,\pi^{3}\,m_{\chi}^{4}}\left(\left|h_{\alpha 1} \right|^{2} \left|h_{\beta 2,3} \right|^{2}+\left|h_{\beta 1} \right|^{2} \left|h_{\alpha 2,3} \right|^{2}\right)\,.
\end{eqnarray}
Therefore we obtain parameter regions allowed by BBN, which is shown in Fig.~\ref{fig:bbn23}.The freeze-in mechanism of DM $N_1$ production is not valid in this scenario due to the largess of the Yukawa couplings.

\begin{figure}[!htbp]
\begin{centering}
\includegraphics[width=.465\textwidth]{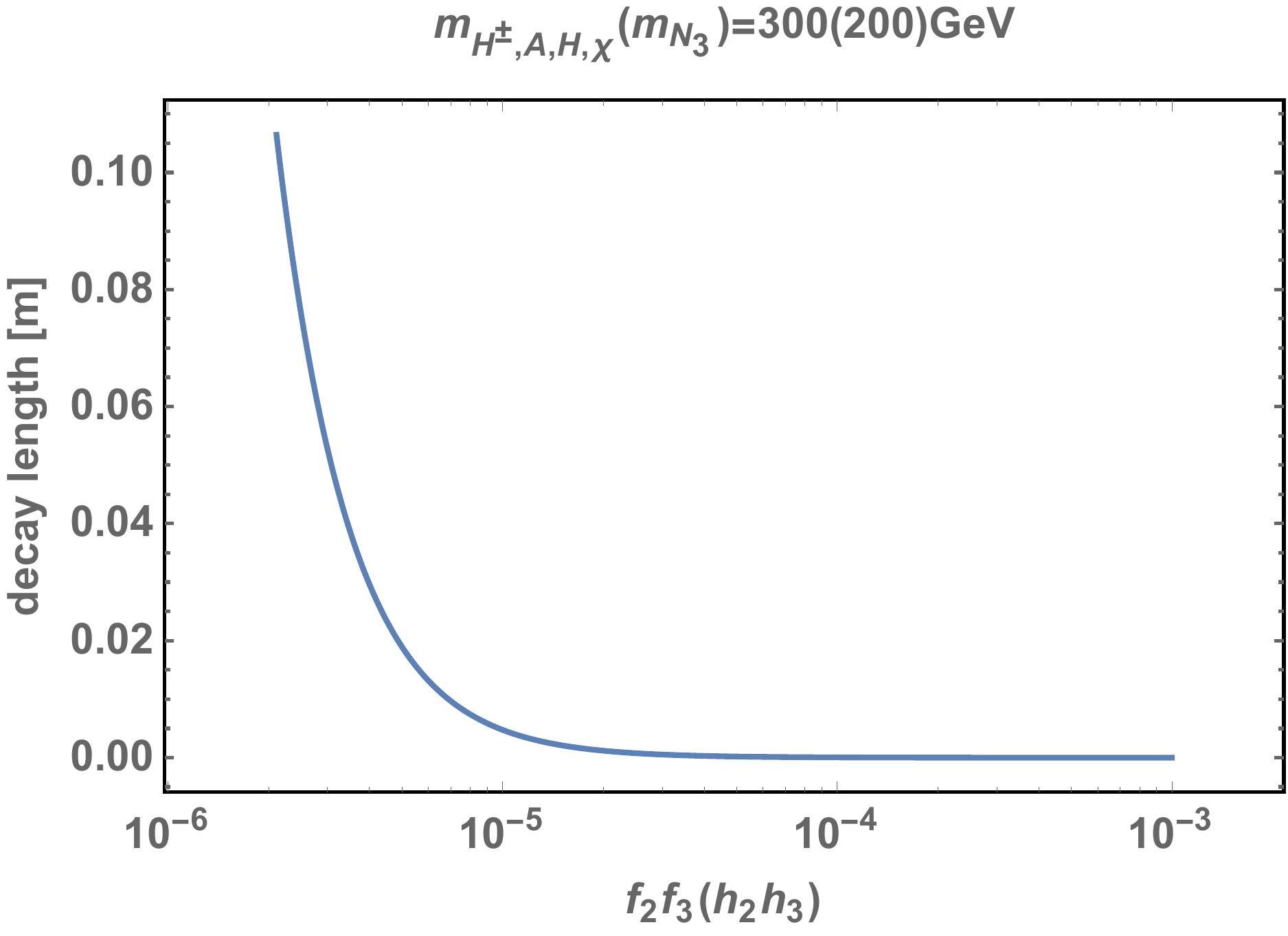}
\caption{The decay length of $N_3$ as a function of $h_{2\alpha}h_{3\beta}(f_{2\alpha}f_{3\beta})$.} \label{fig:dlN3}
\end{centering}
\end{figure}

 Ref.\cite{Hessler:2016kwm} studied the LHC dilepton pair signal in the FIMP DM scenario of the scotogenic model proposed by E.~Ma~\cite{Ma:2006km}, which can be applied to our 
study except the late decay which calls for much heavier $m_{N_{2,3}}$, see Fig.~\ref{fig:N_bpOBC}.
In the model under study, the dilepton decay for $m_{H^\pm,A,H,\chi}>m_{N_{3}}>m_{N_{2}}>m_{N_1}$ is given by
\begin{eqnarray}
	\Gamma'_1(N_{3} \to N_{2}  \,\overline{\ell_{\alpha}}\, \ell_{\beta}) 
	\,&=&\, \frac{m_{N_{3}}^{5}}{6144\,\pi^{3}\,m_{H^\pm,H,A}^{4}}\left(\left|f_{\alpha 2} \right|^{2} \left|f_{\beta 3} \right|^{2}+\left|f_{\beta 3} \right|^{2} \left|f_{\alpha 2} \right|^{2}\right)\,\\
	\Gamma'_2(N_{3} \to N_{2}  \,\overline{\ell_{\alpha}}\, \ell_{\beta}) 
	\,&=&\, \frac{m_{N_{3}}^{5}}{6144\,\pi^{3}\,m_{\chi}^{4}}\left(\left|h_{\alpha 2} \right|^{2} \left|h_{\beta 3} \right|^{2}+\left|h_{\beta 3} \right|^{2} \left|h_{\alpha 2} \right|^{2}\right)\,,
\end{eqnarray}
where we have an additional decay channel of $N_{3} \to N_{2}  \,\overline{\ell_{\alpha}}\, \ell_{\beta} $. The decay length is shown in Fig.~\ref{fig:dlN3}, which means the channel can be tested at the LHC.

\section{Conclusion and discussion}
\label{sec:conl}

This work demonstrate that the two-step SFOEWPT modifies the thermal history of the FIMP DM production through modifying the kinematical thresholds of the decay and scattering processes. 
The phase transition effects may result in a  larger or smaller DM relic density in comparison with the traditional calculation. We study: (1) the FIMP DM production via the the $2\to2$ scattering processes and/or the decay of the particles from the thermal bath; (2) and the DM production with the late decay of a neutral scalar, which is also produced through the freeze-in mechanism. 
The hidden scalars that induce the two-step SFOEWPT would lead to detectable signals at colliders and the gravitational wave signals to be probed on the future space-based interferometer. The realization of the second scenario here requires a tinny mixing angle between the inert CP-even neutral scalar and the singlet scalar to accommodate massive neutrinos of the scotogenic model. 
The SFOEWPT modified FIMP DM is general and apply to the context of DM models with more than one hidden sectors.

For the explanation of the baryon asymmetry of the Universe utilizing the electroweak baryogenesis mechanism~\cite{Morrissey:2012db}, an extra CP-violation operators involving the dark sector are necessary~\cite{ Jiang:2015cwa,Huang:2018aja,Grzadkowski:2018nbc,Vaskonen:2016yiu}.  For the leptogenesis realization in the scotogenic model and it's connection with the WIMPs DM, we refer to Refs.~\cite{Borah:2018rca,Borah:2018uci,Baumholzer:2018sfb,Hugle:2018qbw}.

%=========================================
%\addcontentsline{toc}{section}{Acknowledgments\,}
%\vspace*{2mm}
\section*{Acknowledgments}
\vspace*{-2mm}

We are grateful to Lian-Tao Wang, Stefano Profumo and Michael A. Schmidt for communication on the entropy during the EWPT process. We thank Oscar Zapata, Probir Roy,
and Takashi Toma for discussions on the non-thermal dark matter production mechanisms; and Huai-Ke Guo for helpful communications on the GWs from the SFOEWPT; and Zhi-Long Han for helpful discussions on the scotogenic models. 
The work of LGB is Supported by the National Natural Science Foundation of China (under
grant No.11605016 and No.11647307), Basic Science Research Program through the National Research Foundation of Korea (NRF) funded by the Ministry of Education, Science and Technology (NRF-2016R1A2B4008759), and Korea Research Fellowship Program through the National Research Foundation of Korea (NRF) funded by the Ministry of Science and ICT (2017H1D3A1A01014046).
The work of X.~Liu is supported in part  
by the National Postdoctoral Program for Innovative Talents
under Grant No.\ BX201700116, 
by the Jiangsu Planned Projects for Postdoctoral Research Funds 
under Grant No.\ 1701130B,
by the National Natural Science Foundation of China under Grant Nos.\ 
11775109 and U1738134, 
and by the Nanjing University Grant 14902303.

%=========================================

%\newpage
\vspace*{8mm}
\appendix

\section{Theoretical  and Experimental constraints}
\label{sec:cons}

\subsection{Parameter relations}
\label{sec:pars}
In  the basis of real neutral scalars, $( S, H_0, A )$,
the mass matrix for the dark scalars is
\begin{equation}
\mathbf{M}^2 = \left(\begin{array}{ccc} 2 \mu_S^2 +  v^2
  \lambda_{s\phi} &  v\mu_{\text{soft}} & 0 \\
  v\mu_{\text{soft}}&  \mu_\eta^2 +  v^2  \lambda_L & 0 \\
0 & 0 &  \mu_\eta^2  + v^2 \lambda_S
\end{array} \right).
\label{mix}
\end{equation}

The masses and interactions of the scalar sector are parameterized by the scalar quartic couplings and $\mu_{\Phi,\eta,S,\text{soft}}$,  which are in terms of the physical masses in the global EW vacuum
\begin{eqnarray}
\lambda_3 & = & \frac{2}{v^2}(-m_H^2 \cos^2\theta + m_{H^{\pm}}^2 -m_\chi^2 \sin^2\theta + \lambda_L v^2)\; , \\
\lambda_4 & = & \frac{1}{v^2}(m_{A}^2 + m_H^2 \cos^2\theta - 2 m_{H^{\pm}}^2 + m_\chi^2 \sin^2\theta)\; , \\
\lambda_5 & = & \frac{1}{v^2}(-m_{A}^2 + m_H^2 \cos^2\theta + m_\chi^2 \sin^2\theta)\; , \\
\mu_S^2   & = & \frac{1}{2}(m_\chi^2 \cos^2\theta +m_H^2 \sin^2\theta  -\lambda_{s\phi} v^2)\; , \\
\mu_\eta^2   & = &  m_H^2 \cos^2\theta + m_\chi^2 \sin^2\theta -\lambda_L v^2\; , \\
\mu_{\text{soft}} & = &\frac{1}{2v}\left(m_{\chi}^2-m_{H}^2\right)\sin2\theta\; ,
\label{eq:scalarcouplings}
\end{eqnarray}
with $\lambda_{L,S}=(\lambda_3+\lambda_4\pm\lambda_5)/2$.

\subsection{Perturbative unitarity}
\label{sec:unitarity}

%In the high energy limit, the quartic contact interaction terms contribute to the tree-level scalar-scalar scattering matrix dominantly. The $s$-wave scattering amplitudes are constrained by the perturbative unitarity limits, which requires that the eigenvalues of the $S$-matrix $\mathcal{M}$ must be smaller than the unitarity bound:
%\begin{equation}
%  \abs{\Re \mathcal{M}} < \frac{1}{2}.
%\end{equation}
%
%The perturbative unitarity of the general two-Higgs-Doublet model were first studied in Ref.~\cite{Kanemura:1993hm,Akeroyd:2000wc}. In this work we extend the formalism of Ref.~\cite{Ginzburg:2003fe,Ginzburg:2005dt} for the IDM  to states containing an extra singlet $S$. The initial states are classified according to their total hypercharge $Y$ ($0$, $1$ or $2$), weak isospin $\sigma$ ($0$, $\frac{1}{2}$ or $1$) and discrete $\Z_2$ charge $X$. 
%
%For simplicity, we only list here the initial states with hypercharge $Y = 0$ and $\sigma = 0$ which differ from the 2HDM initial states\cite{Ginzburg:2003fe,Ginzburg:2005dt}:
%%The full set of possible initial states with hypercharge $Y = 1$ and $\sigma = \frac{1}{2}$ is
%%\begin{equation}
%%\Phi_{H} S,  \Phi_{I} S.
%%\end{equation}
%\begin{equation}
% \frac{1}{\sqrt{2}} \hc{\Phi_{H}} \Phi_{H}, 
% \frac{1}{\sqrt{2}} \hc{\Phi_{I}} \Phi_{I}, 
% \frac{1}{\sqrt{2}} S^{2}, 
% \frac{1}{\sqrt{2}} \hc{\Phi_{H}} \Phi_{I}, 
% \frac{1}{\sqrt{2}} \hc{\Phi_{I}} \Phi_{H},
%\end{equation}
%where the first three states are even under $\Z_{2}$ and the last two states are odd.
%In the following, we present 

All the scattering matrices of the model are~\cite{Liu:2017gfg}
%Firstly, the scattering matrices are given as
\begin{align}
  8 \pi S_{Y=2,\sigma=1} &= 
  \begin{pmatrix}
    2 \lambda_{1} & \lambda_{5} & 0 \\
    \lambda_{5}^{*} & 2 \lambda_{2} & 0 \\
    0 & 0 & \lambda_{3} + \lambda_{4}
  \end{pmatrix},
  &
  8 \pi S_{Y=2,\sigma=0} &= \lambda_{3} - \lambda_{4},
  \\
  8 \pi S_{Y=0,\sigma=1} &= 
  \begin{pmatrix}
    2 \lambda_{1} & \lambda_{4} & 0 & 0 \\
    \lambda_{4} & 2 \lambda_{2} & 0 & 0 \\
    0 & 0 & \lambda_{3} & \lambda_{5}^{*} \\
    0 & 0 & \lambda_{5} & \lambda_{3}
  \end{pmatrix},
  &
  8 \pi S_{Y=1,\sigma=1/2} &= 
    \begin{pmatrix}
      2 \lambda_{s\phi} & 0  \\
      0 & 2 \lambda_{s\eta}    
    \end{pmatrix},
\end{align}
\begin{equation}
    8 \pi S_{Y=0,\sigma=0} = 
  \begin{pmatrix}
    6 \lambda_{1} & 2 \lambda_{3} + \lambda_{4} & \sqrt{2} \lambda_{s\phi} & 0 & 0  \\
    2 \lambda_{3} + \lambda_{4} & 6 \lambda_{2} & \sqrt{2} \lambda_{s\eta} & 0 & 0  \\
    \sqrt{2} \lambda_{s\phi} & \sqrt{2} \lambda_{s\eta} & \lambda_{S}/4 & 0 & 0  \\
      0 & 0 & 0  & \lambda_{3} + 2 \lambda_{4} & 3 \lambda_{5}^{*}   
    \\
    0 & 0 & 0 & 3 \lambda_{5} & \lambda_{3} + 2 \lambda_{4} \\
  \end{pmatrix}.
\end{equation}

Then the eigenvalues $\Lambda_{Y\sigma i}^{X}$ of the above scattering matrices (where $i = \pm \text{ or }1,2,3$) can be calculated as
\begin{align}
%\begin{split}
  \Lambda_{21\pm}^{even} &= \lambda_{1} + \lambda_{2} \pm \sqrt{(\lambda_{1} - \lambda_{2})^{2} + \abs{\lambda_{5}}^{2}},~
  \Lambda_{21}^{odd} = \lambda_{3} + \lambda_{4}\;,~
  \Lambda_{20}^{odd} = \lambda_{3} - \lambda_{4}\; ,  \\
   \Lambda_{01\pm}^{even} &= \lambda_{1} + \lambda_{2} \pm \sqrt{(\lambda_{1} - \lambda_{2})^{2} + \lambda_{4}^{2}}\;,\qquad
  \Lambda_{01\pm}^{odd} = \lambda_{3} \pm \abs{\lambda_{5}},
  \\
    \Lambda_{00\pm}^{odd} &= \lambda_{3} +2\lambda_{4}\pm 3\abs{\lambda_{5}},
%  \\
%  {\color{red}\abs{ \Lambda_{00\, 1,2,3}^{even} }} &{\color{red}\leqslant \frac{1}{3} \left( 6 \lambda_1 + 6 \lambda_2 + \lambda_S + 2 \, [ 36 (\lambda_{1}^{2} - \lambda_{1} \lambda_{2} + \lambda_{2}^{2}) \phantom{^{\frac{1}{2}}} 
% \right.} \\
% &{\color{red} \left. - 6 (\lambda_{1} + \lambda_{2}) \lambda_{6} + \lambda_{6}^{2} 
% + 3 (2 \lambda_{3} + \lambda_{4})^{2} + 6 (\lambda_{7}^{2} + \lambda_{8}^{2}) ]^{\frac{1}{2}} \right)}, \notag
%\end{split}
\label{eq:Z2:unitarity}
\end{align}
and the $\Lambda_{00\, 1,2,3}^{\text{even}}$ correspond to  % given by $\frac{1}{16 \pi} \times$ 
the three roots of the polynomial equation
\begin{equation}
\begin{split}
  & x^3 - x^2 (6 \lambda_{1} + 6 \lambda_{2} +  \lambda_{s}/4) + x (36\lambda_{1} \lambda_{2}-4\lambda_{3}^2-4\lambda_{3} \lambda_{4}- \lambda_{4}^2+3\lambda_{1} \lambda_{s}/2+3\lambda_{2} \lambda_{s}/2 \\
   &- 2\lambda_{s\phi}^2 - 2\lambda_{s\eta}^2) 
   +12 \lambda_{1} \lambda_{s\eta}^2 +
 12 \lambda_{2} \lambda_{s\phi}^2 - 4\lambda_{4}\lambda_{s\phi}\lambda_{s\eta}- 8\lambda_{3}\lambda_{s\phi}\lambda_{s\eta}+\lambda_{4}^2\lambda_{s}/4+\lambda_{3}\lambda_{4}\lambda_{s}\\
 &+\lambda_{3}^2\lambda_{s}-9\lambda_{1}\lambda_{2}\lambda_{s}
 = 0.
\end{split}
\end{equation}

%{\color{red}Since $\Lambda_{00\, 1,2,3}^{even}$ are too cumbersome to be presented in full, in \eqref{eq:Lambda000} we have given an upper bound on their absolute values by applying Samuelson's inequality \cite{samuelson} to the characteristic equation (in our numerical calculations we use the exact eigenvalues). }

\subsection{Vacuum stability conditions }

%And following the approach 
The vacuum stability conditions for the model are given as~\cite{Liu:2017gfg},
\begin{align}
  \lambda_{1} \; ,  ~\lambda_{2}\;  ,~ \lambda_{s} > 0\; , ~~~~~~~~~
 \sqrt{\lambda_{1} \lambda_{s}} + \lambda_{s\phi} > 0 ,~~~~~~~~~
   \sqrt{\lambda_{2} \lambda_{s}} + \lambda_{s\eta} > 0\; , 
  \end{align}
  and for $\lambda_4>0$ case,
  \begin{align}
&2 \sqrt{\lambda_{1} \lambda_{2}} + \lambda_{3} > 0\; ,\\ 
   & \sqrt{\lambda_{1} \lambda_{2} \lambda_{s}}/2 
  + \sqrt{\lambda_{1}} \lambda_{s\eta} + \sqrt{\lambda_{2}} \lambda_{s\phi}
  + \sqrt{\lambda_{s}} \lambda_{3}/2\notag \\
  &+ \sqrt{(\sqrt{\lambda_{1} \lambda_{s}} + \lambda_{s\phi}) (2 \sqrt{\lambda_{2} \lambda_{s\phi}} + \lambda_{s\eta})
  (2 \sqrt{\lambda_{1} \lambda_{2}} + \lambda_{3})} > 0\; ,
    \end{align}
  if $\lambda_4-|\lambda_5|<0$, one obtains~\cite{Kannike:2012pe},  
  \begin{align}
  &2 \sqrt{\lambda_{1} \lambda_{2}} + \lambda_{3} + \lambda_{4} - |\lambda_{5}| > 0\; ,\label{eq:idm} \\
  &   \sqrt{\lambda_{1} \lambda_{2} \lambda_{s}}/2 
  + \sqrt{\lambda_{1}} \lambda_{s\eta} + \sqrt{\lambda_{2}} \lambda_{s\phi}
  + \sqrt{\lambda_{s}} (\lambda_{3} + \lambda_{4} - |\lambda_{5}|)/2 \notag\\
  &+ \sqrt{(\sqrt{\lambda_{1} \lambda_{s}} + \lambda_{s\phi}) (2 \sqrt{\lambda_{2} \lambda_{s}} + \lambda_{s\eta})
  (2 \sqrt{\lambda_{1} \lambda_{2}} + \lambda_{3} + \lambda_{4} - |\lambda_{5}|)} > 0\; .
\end{align}

\subsection{The $T$ parameter}

The $T$ parameter is given by $T=\Delta\rho/\alpha_{EM}$~\cite{Barbieri:2006dq}, where~\cite{Liu:2017gfg} 
\begin{align}\label{rho1}
\Delta\rho &  =\frac{(\lambda_{4}+\lambda_{5})^{2}}{2}\left(\cos^2\theta f(m_{_{H^\pm}},m_{H})+\sin^2\theta f(m_{_{H^\pm}},m_{\chi})\right)\nonumber\\
&+\frac{(\lambda
_{4}-\lambda_{5})^{2}}{2}f(m_{_{H^{\pm}}},m_{A})-2(\lambda_{5}\cos\theta-\mu_{\text{soft}}\sin\theta/v)^2 f(m_{A},m_{H})\nonumber\\
&-2(\lambda_{5}\sin\theta+\mu_{\text{soft}}\cos\theta/v)^2f(m_{A},m_{\chi})\; ,\\
\nonumber\\
\mathrm{with}&~~~ f(m_{1},m_{2})    =\frac{v^{2}}{32\pi^{2}}\int_{0}^{1}\frac{dx\,x(1-x)}%
{xm_{1}^{2}+(1-x)m_{2}^{2}}\; ,
\end{align}
$\mu_{\text{soft}}$ and $\lambda_{4,5}$ are given in Eq.~\ref{eq:scalarcouplings}.

\subsection{LEP bounds}

Considering the CP-even neutral Higgs $H$ is a mixture of doublet and singlet, the LEP bounds on the scalar masses of this model are the same as in IDM. The lower limits on the  scalar masses comes from the precise measurements of the $W$ and $Z$ widths,
\begin{eqnarray}
\label{eq:constr-widths}
&& m_{H} + m_{H^{\pm}} > m_{W^{\pm}}\;, \quad \quad m_{A} + m_{H^{\pm}} > m_{W^{\pm}} \; ,\nonumber\\
&& m_{H} + m_{A} > m_{Z}\; , \quad  \quad 2m_{H^{\pm}} > m_{Z}\; ,
\end{eqnarray}
Utilizing the neutralino search at LEP II, the mass splitting $\Delta m_{HA}$ should be either smaller than 8 GeV or greater than 100 GeV for $m_A < 80$ GeV~\cite{Lundstrom:2008ai}.
The production of the charged Higgs pairs $H^+H^-$ at the LEPII give rise to~\cite{Pierce:2007ut},
\begin{equation}
m_{H^\pm}>70\mbox{ GeV}\; .
\label{eq:mhcp-lep2}
\end{equation}

\subsection{On the massive neutrino mass and the constraints on the model parameters }

\begin{figure}[!htp]
\begin{centering}
\includegraphics[width=.3\textwidth]{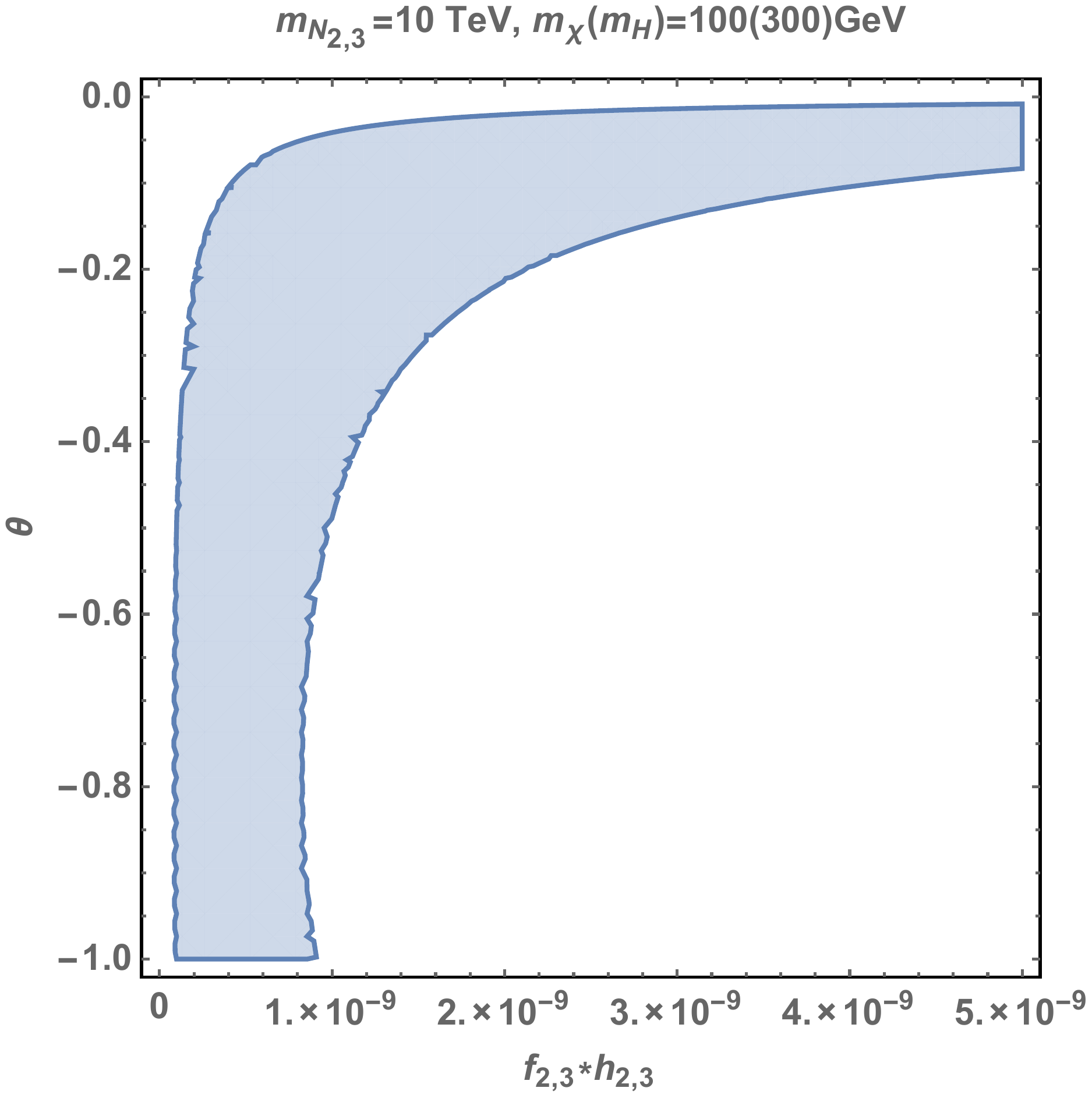}
\includegraphics[width=.3\textwidth]{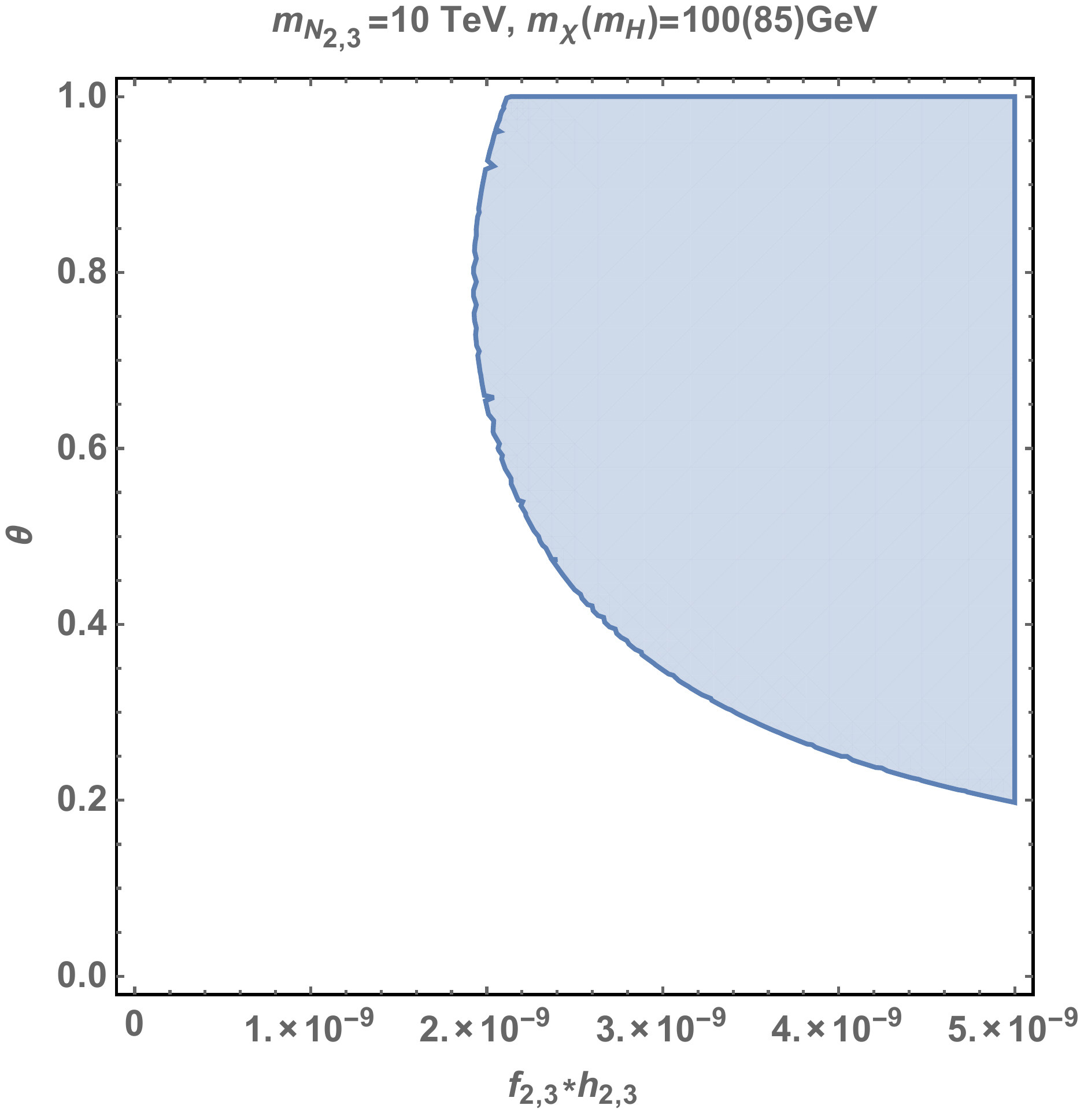}
\includegraphics[width=.3\textwidth]{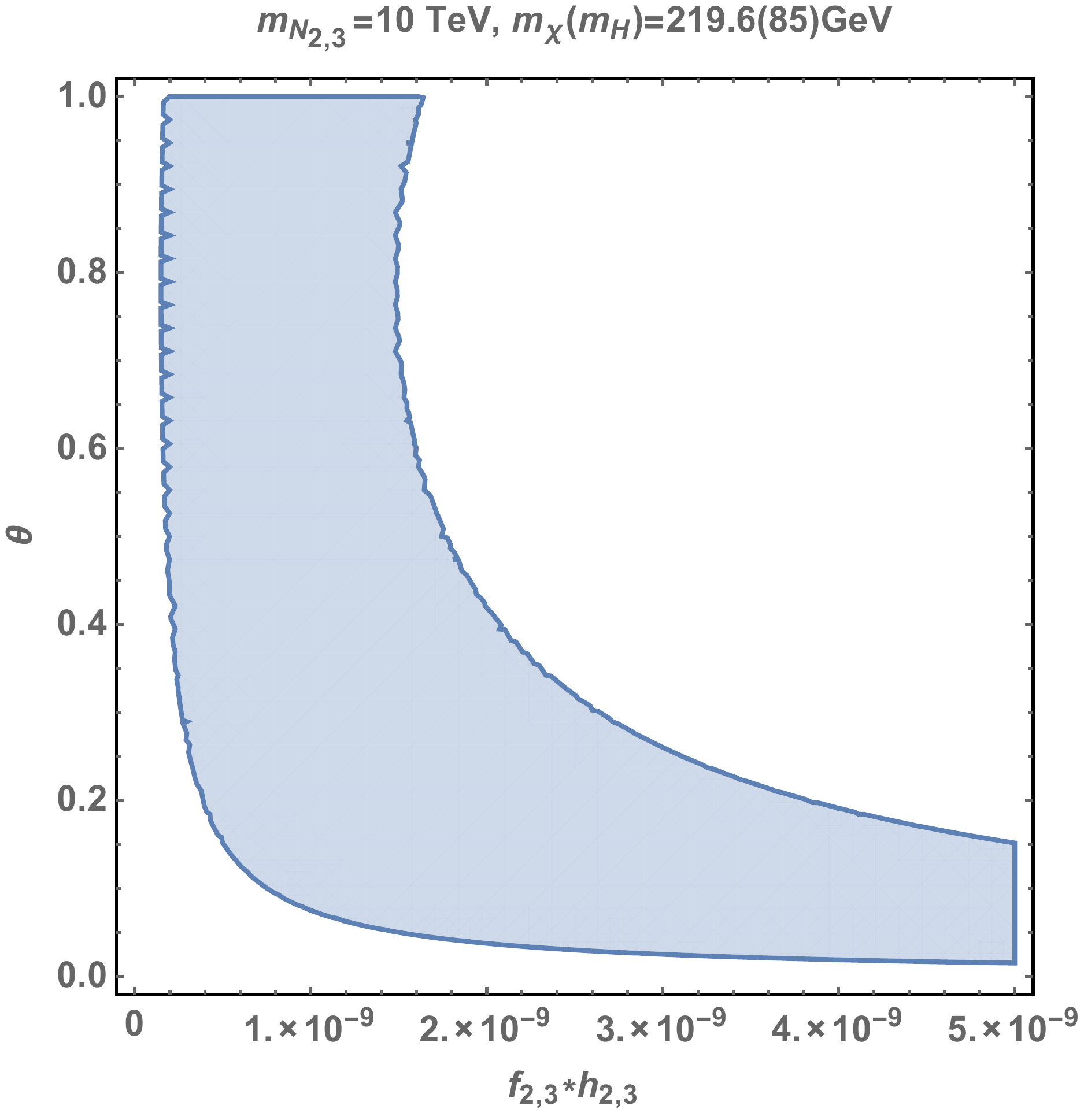}
\caption{The neutrino mass bounds on mixing angle $\theta$ and the multiplication of Yukawa $f$ and $h$ for ${\rm OAC}$(left), ${\rm OBC}$ (middle), ${\rm OBC}$ with late decay (right).} \label{fig:MN2}
\end{centering}
\end{figure}

For large mass splitting of $m_\chi$ and $m_H$ and large $m_{N_{2,3}}$, Eq.~\ref{eq:neumM} reduces to
\begin{eqnarray}
\left(M_\nu\right)_{\alpha\beta}=\frac{\sin2\theta}{32\sqrt{2}\pi^2}\sum_{\kappa=2,3} f_{\alpha \kappa}h_{\kappa\beta} m_{N_\kappa}\Big[\frac{m_H^2}{m_{N_\kappa}^2}\log \frac{m_H^2}{m_{N_\kappa}^2}-\frac{m_\chi^2}{m_{N_\kappa}^2}\log \frac{m_\chi^2}{m_{N_\kappa}^2}\Big]\;,
\end{eqnarray}
The corresponding parameter limits from the neutrino mass are plotted in Fig.~\ref{fig:MN2}, a larger mixing angle $\theta$ is preferred for a lower magnitude of $f_{\alpha2,3 }h_{2,3\beta}$.

Considering the $N_1$ decoupled from the neutrino mass matrix, when $m_\chi\sim m_H$ the neutrino mass matrix of Eq.~\ref{eq:neumM} reduces to
\begin{eqnarray}
\left(M_\nu\right)_{\alpha\beta}=\frac{\sin2\theta}{32\sqrt{2}\pi^2}\sum_{\kappa=2,3} f_{\alpha \kappa}h_{\kappa\beta} \frac{m_{N_\kappa}(m_H^2-m_\chi^2)}{m_{H\chi}^2-m_{N_\kappa}^2} \Bigg(1- \frac{m_{N_\kappa}^2}{m_{H\chi}^2-m_{N_\kappa}^2}\log\left(\frac{m_{H\chi}^2}{m_{N_\kappa}^2}\right)\Bigg)\;,\label{eq:simHchi}
\end{eqnarray}
with $m_{H\chi}^2=(m_H^2+m_\chi^2)/2$. With Eq.~\ref{eq:scalarcouplings}, for the parameter set of $m_{H\chi}^2\ll m_{N_\kappa}^2$, the equation Eq.~\ref{eq:simHchi} casts the form of,
 \begin{eqnarray}\label{eq:mnheavy}
\left(M_\nu\right)_{\alpha\beta}&=&\frac{\sin2\theta}{32\sqrt{2}\pi^2}\sum_{\kappa=2,3} f_{\alpha \kappa}h_{\kappa\beta} \frac{(m_H^2-m_\chi^2)}{m_{N_\kappa}} \left(\log\frac{m_{N_\kappa}^2}{m_{H\chi}^2}-1\right)\;\nonumber\\
&=&\frac{1}{32\sqrt{2}\pi^2}\sum_\kappa f_{\alpha \kappa}h_{\kappa\beta} \frac{-2v\mu_{soft}}{m_{N_\kappa}} \left(\log\frac{m_{N_\kappa}^2}{m_{H\chi}^2}-1\right)\;,
\end{eqnarray}
whereas, when $m_{H\chi}^2\gg m_{N_\kappa}^2$, the Eq.~\ref{eq:simHchi} recasts the form of
  \begin{eqnarray}
  \label{eq:mnlight}
\left(M_\nu\right)_{\alpha\beta}&=&\frac{\sin2\theta}{32\sqrt{2}\pi^2}\sum_{\kappa=2,3} f_{\alpha \kappa}h_{\kappa\beta} \frac{(m_H^2-m_\chi^2)m_{N_\kappa}}{m_{H\chi}^2}\;,\nonumber\\
&=&\frac{-2v\mu_{soft}}{32\sqrt{2}\pi^2}\sum_\kappa f_{\alpha \kappa}h_{\kappa\beta} \frac{m_{N_\kappa}}{m_{H\chi}^2}\;.
\end{eqnarray}
Since we assume the $N_1$ never reachs equilibrium with the thermal bath, the requirement of the neutrino mass of order $\sim 0.1$ eV 
leads to a ballpark estimate of the size of the multiplication of Yukawa couplings
 $f_{\alpha2,3 }h_{2,3\beta}$ and the soft terms $\mu_{soft}$.

%%%%%%%%%%%%%%%%%%%

\section{Thermal field dependent masses}
\label{app:thermalmass}

The field dependent mass of CP-even, CP-odd, and the charged scalars are given as follows,
\begin{eqnarray}
M_h^2&=&\left(
\begin{array}{ccc}
 3 \lambda_1 h^2+ \lambda_L H_0^2+\lambda_{s\phi } S^2 +\mu_\phi^2 &  2\lambda_L h H_0 +\mu_{soft} S   & 2\lambda_{s\phi} h S +\mu_{soft} H_0 \\
  2\lambda_L h H_0 +\mu_{soft} S   &  \lambda_L  h^2+3  \lambda_2 H_0^2+\lambda_{s\eta} S^2 +\mu_\eta^2 & 2 \lambda_{s\eta} H_0 S +\mu_{soft} h  \\
 2\lambda_{s\phi} h S +\mu_{soft} H_0 & 2\lambda_{s\eta} H_0 S +\mu_{soft} h & \lambda_{s\phi} h^2+3\lambda_s S^2 + \lambda_{s\eta} H_0^2 +\mu_s^2   \\
\end{array}
\right)\;,\nonumber\\
M_A^2&=&
\left(
\begin{array}{cc}
\lambda _1 h^2+ \lambda_L H_0^2 +  \lambda_{s\phi} S^2+\mu_\phi^2 &   \lambda_5 h H_0+\mu_{soft} S  \\
  \lambda_5 h H_0+\mu_{soft} S  & \lambda_S h^2+\lambda_2 H_0^2 + \lambda_{s\eta} S^2 +\mu_\eta^2 \\
\end{array}
\right)\;,\nonumber\\
M_{H^\pm}^2&=&
\left(
\begin{array}{cc}
\lambda_1 h^2+\frac{ \lambda_3}{2}H_0^2+ \lambda_{s\phi} S^2+\mu_\phi^2 & \frac{ \lambda_4+\lambda_5}{2} h H_0+ \mu_{soft} S \\
 \frac{ \lambda_4+\lambda_5}{2} h H_0+ \mu_{soft} S  & \frac{\lambda_3 }{2}h^2+ \lambda_2 H_0^2+ \lambda_{s\eta} S^2 +\mu_\eta^2 \\
\end{array}
\right)\;.
\end{eqnarray}

The field dependent thermal masses of scalar masses can be obtained by the replacement of $\mu_i^2\rightarrow \mu_i^2( T) $ being given in Eq.~\ref{eq:thmu}.
As the temperature cools down, the thermal masses of the neutral ($m_{\chi,H,A}(T)$) and charged Higgs bosons ($m_{H^\pm}(T)$) for the study of thermally modified FIMP is different in the symmetric phase, the $Z_2$ broken phase, and the EW symmetry broken phases. In the symmetric phase, one have,
\begin{eqnarray}
M_h^{sys}(T)&=&M_h(T)|_{\langle h,H_0,S\rangle \to 0}\;,\\
M_A^{sys}(T)&=&M_A(T)|_{\langle h,H_0,S\rangle\to 0}\;,\\
M_{H^\pm}^{sys}(T)&=&M_{H^\pm}|_{\langle h,H_0,S\rangle\to 0}\;.
\end{eqnarray}
In the $Z_2$ broken phase, we have
\begin{eqnarray}
M_h^{\slashed{Z_2}}(T)&=&M_h(T)|_{\langle h\rangle \to 0, \langle H_0 {\rm ~or~} S\rangle \to v_H(T) {\rm~ or~} v_s(T) }\;,\\
M_A^{\slashed{Z_2}}(T)&=&M_A(T)|_{\langle h\rangle \to 0, \langle H_0 {\rm ~or~} S\rangle \to v_H(T) {\rm ~or~} v_s(T) }\;,\\
M_{H^\pm}^{\slashed{Z_2}}(T)&=&M_{H^\pm}|_{\langle h\rangle \to 0, \langle H_0 {\rm ~or~} S\rangle \to v_H(T) {\rm ~or~} v_s(T) }\;,
\end{eqnarray}
depending on if the first-stage second order phase transition is the inert phase or the singlet phase.
Finally, in the EW broken phase, we have,
\begin{eqnarray}
M_h^{\slashed{EW}}(T)&=&M_h(T)|_{\langle h\rangle \to v(T), \langle H_0 , S\rangle \to 0}\;,\\
M_A^{\slashed{ EW}}(T)&=&M_A(T)|_{\langle h\rangle \to v(T), \langle H _0,S\rangle \to 0 }\;,\\
M_{H^\pm}^{\slashed{EW}}(T)&=&M_{H^\pm}|_{\langle h\rangle \to v(T), \langle H_0,S\rangle \to 0 }\;.
\end{eqnarray}

%%%%%%%%%%%%%%%%%%%%
\section{The DM production process}
\label{app:xec}

For the phase transition pattern $O\to A\to C$, the FIMP DM $N_1$ production process is dominated by the decay of $X\to N\ell/\nu$ with $X=\chi,H,A,H^\pm$ living in the thermal equilibrium. For the phase transition pattern $O\to B \to C$, both decay ($H,A,H^\pm\to N\ell$) and $2\to 2$ scattering process ($\gamma H^\pm \to \bar{N}\ell^\pm$) can contribute to the DM production, with the latter dominate.   
In this scenario, the late decay of $\chi \to N_1 \nu $ after freeze in of $\chi$ may contribute to the DM abundance of $N_1$ dominantly, with the subdominant contribution coming from the $2\to 2$ scattering process ($\gamma H^\pm \to \bar{N}\ell^\pm$).  

\begin{figure}[!htbp]
\begin{centering}
\includegraphics[width=.3\textwidth]{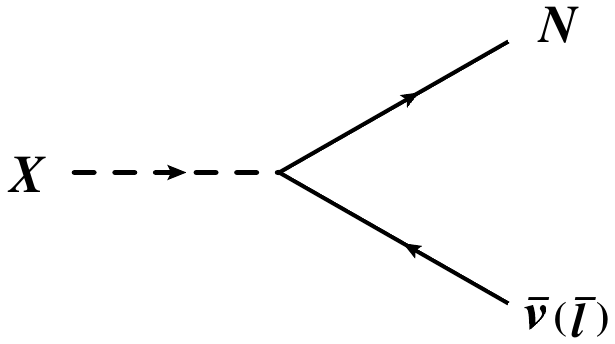}
\includegraphics[width=.3\textwidth]{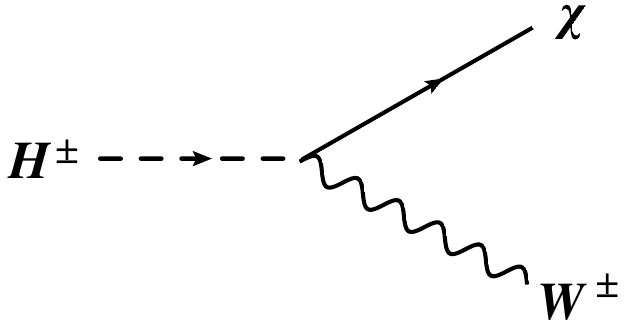}\\
\includegraphics[width=.6\textwidth]{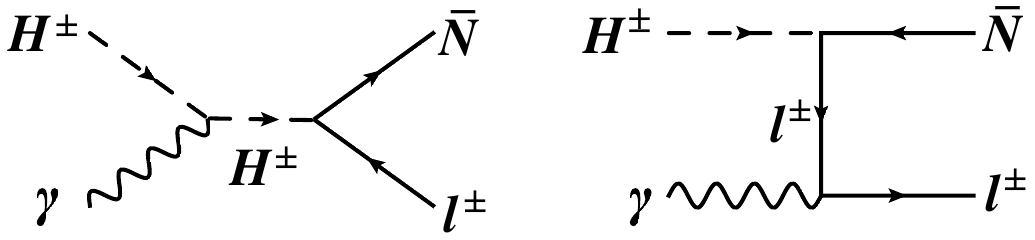}
\caption{The DM production process.} \label{fig:DMpro}
\end{centering}
\end{figure}

With the amplitude of the scattering process being calculated by the {\tt CalcHEP}~\cite{Belyaev:2012qa} after the model being implemented in the {\tt FeynRules}~\cite{Alloul:2013bka}, 
\begin{eqnarray}
\abs{\mathcal{M}}^2&=&4 \pi  a_{EW} f_N^2\Big (\frac{s \left(2(m_\ell^2+ M_N^2)- s\right)-M_{H^\pm}^4}{\left(M_{H^\pm}^2-s\right)^2}+\frac{2m_\ell^2(m_\ell^2+m_N^2-m_{H^\pm}^2)}{(m_\ell^2-t)^2}\nonumber\\
&&+\frac{-2m_\ell^4+2 m_\ell^2 (M_{H^\pm}^2-s)+M_{H^\pm}^4-2M_{H^\pm}^2 M_N^2+2 M_N^4-2M_N^2 s+s^2}{(M_{H^\pm}^2 -s)( t-m_\ell^2)}\Big)\;,\nonumber
\end{eqnarray}
with $t=(p_1(\gamma)-p_3(\ell))^2$ and $s=(p_1(\gamma)+p_2(H^\pm))^2$, 
the differential cross section is given by
\begin{eqnarray}
&&d\sigma _{\gamma H^\pm\to \ell N}/d\Omega=\frac{1}{64\pi^2}\frac{\sqrt{1-4 m_N^2/s}}{s\sqrt{1-4m_{H^\pm}^2/s}}\abs{\mathcal{M}}^2\;,
\end{eqnarray}
With the Mandelstam variable $t$ being, 
\begin{eqnarray}
t&=& m_\ell^2-2 p_1(\gamma)\cdot p_3(\ell)\nonumber\\
&\equiv& m_\ell^2+\frac{(m_{H^\pm}^2-s)}{2s}\left(m_\ell^2 -\cos\theta\sqrt{m_\ell^4-2 m_\ell^2(s+m_N^2)+(s-m_N^2)^2} +s-m_N^2\right)\;, 
\end{eqnarray}
 one can obtain the Dirac neutrino $N_1$ production cross section after integrated out the collide angle $\theta$,
 \begin{eqnarray}
 &&\sigma _{\gamma H^\pm\to \ell N}=\frac{1}{32\pi} \frac{\sqrt{1-4 m_N^2/s}}{s\sqrt{1-4m_{H^\pm}^2/s}}\int d\cos\theta|\mathcal{M}|^2
 \end{eqnarray}

%%%%%%%%%%%%%%%%%%%%%

\section{Entropy injected by EWPT}
\label{app:entropy}

In this section, we recall knowledge of entropy deviation induced by EWPT~\cite{Wainwright:2009mq}.
Assuming that reheating
happens quickly relative to the expansion rate, the energy density $\rho$ of the universe does 
not change during reheating. And there are only small amount of reheating by the release latent heat of the transition. We do not expect the phase coexistence stage for SFOEWPT.

The injection of entropy from  by supercooling process of EWPT for the two step pattern is evaluated by~\cite{Patel:2012pi}, 
\begin{equation}
\Delta s=-(\frac{dV}{dT}|_\phi-\frac{dV}{dT}|_H)
\end{equation}
with the finite temperature potential $V$ included only the thermal mass correction evaluated at the nucleation temperatures, which has been normalized by the SM entropy in the radiation dominate universe, 
\begin{equation}
s=\frac{2\pi^2}{45}g_{\star s} T_{EW}^3\;.
\end{equation}
By taking $g_{\star s}=100$ there, it was estimated that $\Delta s/s_{EW}$ is around percent level around the phase transition temperature $T_n$ for benchmarks of $O\to A(B)\to C$.
As in the Standard model,
(see Ref.~\cite{Quiros:1999jp}),  one finds a negligible dilution factor, namely:
\begin{equation}
\left(\frac{a_f}{a_i}\right)^3 = \frac{s_+}{s_-} \approx 1 + \frac{\Delta s}{s_+}\approx 1.01,
\end{equation}
with the entropy difference between the high temperature and low temperature entropy: $\Delta s= s_+-s_-$.

\bibliographystyle{JHEP}

\end{document}